\newcommand{\logg}{$\log\,g$}
\newcommand{\teff}{$T_{\rm eff}$}

\documentclass{aa}
\usepackage{txfonts}
\usepackage{graphicx}
\begin{document}

\title{New Mn II energy levels from the STIS-HST spectrum of the HgMn 
star HD\,175640{\thanks{Tables A.1 and A.2 are only available in electronic
form at CDS via anonymous ftp to cdsarc.u-strasbg.fr (130.79.128.5) or via
http://cdsweb.u-strasbg.fr/cgi-bin/qcat?J/A+A/}} 
}

\author{
F.\, Castelli
\inst{1}
\and
R. L.\,Kurucz
\inst{2}
\and
C. R. \,Cowley
\inst{3}
}

\offprints{F.\,Castelli}

\institute{
Istituto Nazionale di Astrofisica,
Osservatorio Astronomico di Trieste, Via Tiepolo 11,
I-34143 Trieste, Italy\\
\email{castelli@oats.inaf.it}
\and
Harvard-Smithsonian Center for Astrophysics, 60 Garden Street, Cambridge, 
MA 02138,USA\\
\email{rkurucz@cfa.harvard.edu} 
\and
Department of Astronomy, University of Michigan, Ann Arbor, MI 48109-1042, USA\\\email{cowley@umich.edu}
}

\date{}

\abstract
{}
{The NIST database lists several \ion{Mn}{ii} lines that 
were observed in the laboratory but not classified.
They  cannot be used in spectrum synthesis because their atomic line
data are unknown. These lines are concentrated in the 2380-2700\,\AA\ interval.
We aimed to assign energy levels and $\log\,gf$ values to these lines.
}
{Semiempirical line data for \ion{Mn}{ii} computed by Kurucz were used to 
synthesize the ultraviolet spectrum of the slow-rotating, HgMn star 
HD\,175640. The spectrum was compared with the high-resolution  spectrum 
observed with the HST-STIS equipment. A UVES spectrum covering the
3050-10000\,\AA\ region was also examined.
}
{We determined a total of 73 new energy levels, 58 from the STIS spectrum of
HD\,175640 and another 15  from the UVES spectrum. The new energy levels 
give rise to numerous new computed lines.
We have identified more than 50\% of the unclassified lines listed in the
NIST database and have changed the assignment of another 24 lines.
 An abundance analysis of the star HD\,175640, based on
the comparison of observed and computed ultraviolet spectra in the 1250-3040\,\AA\ interval, is
the by-product of this study on \ion{Mn}{ii}. 
}
{}


\maketitle{}

\section{Introduction}

The availability of a reliable and complete set of atomic data is an essential 
requirement for research in stellar physics,
in particular for computing stellar atmospheres
and stellar spectra to be compared with observations
at  various resolutions.

The high$-$resolution, high signal$-$to$-$noise stellar spectra observed 
by the modern telescopes and spectrographs have proved to be a useful 
 tool for studies aimed at extending the knowledge of atomic and molecular data. 
An example is the analysis of the UVES spectrum of the peculiar star HR\,6000 
which has permitted us to fix about 120 \ion{Fe}{ii} energy levels not 
observed 
before in the laboratory, but predicted  with approximate 
energy values by the theory (Castelli et al. 2008, Castelli et al. 2009, 
Castelli \& Kurucz 2010).
Another example of the use of stellar spectra to derive atomic data 
is the paper by Peterson \& Kurucz (2015) who obtained 66 new energy 
levels for \ion{Fe}{i} from the analysis of HST-STIS spectra 
of a group of 13 metal-poor stars. 
In the above works, for each new determined energy level,
several new transitions with wavelengths and oscillator
strengths are associated,
so that the line lists used for computing spectra are considerably augmented.

In this paper we extend to \ion{Mn}{ii} the above described studies.
 History has shown that
the refinement and extension of our atomic database has lead to important
astronomical discoveries. In the case of the \ion{Mn}{ii} spectrum, 
the subject of the present paper, the extended analysis 
by Iglesias \& Velasco (1964) enabled many new line identifications 
in the class of stars now known as mercury-manganese (HgMn) stars,
occupying the spectral region $\sim$ A0/B9-B6 (10500-16000K). This in turn
enabled Bidelman (1962) and Dworetsky (1969) to show that these stars were
enormously enriched in mercury and platinum. In the present study, we have been
able to obtain abundances for ten elements that were unavailable from
the ground spectra (\ion{B}, \ion{N}, \ion{Al}, \ion{Cl}, \ion{V}, 
\ion{Zn}, \ion{Ge}, \ion{As}, \ion{Ag}, and \ion{Cd}). The ability
to distinguish their spectral lines from those of \ion{Mn}{ii} was essential.

To study the \ion{Mn}{ii} spectrum, we adopted the same procedure described 
by Castelli \& Kurucz (2010). We analysed HST-STIS spectra 
from 1250-3045\,\AA\ and a UVES spectrum from 
3050 to 10000\,\AA\ of the HgMn star HD\,175640.
It is a slowly rotating star with an abundance in manganese on the order
of 2.5\,dex over the solar value. The UVES spectrum,
previously analyzed to derive stellar parameters and abundances by
Castelli \& Hubrig (2004), was further examined for this paper to derive new 
\ion{Mn}{ii} energy levels.

To fix new energy levels we adopted the semiempirical 
\ion{Mn}{ii} line data computed by Kurucz. We determined a total of 73
new energy levels, 58  from the HST-STIS spectrum and 15 from the UVES 
spectrum. 
Updated line lists for most elements were  adopted in order 
to synthesize the spectra. 
The most recent work on \ion{Mn}{ii} is that of Kramida \& Sansonetti (2013)
who critically analyzed all the atomic data available in the 
literature for this ion.
They revised the Sugar \& Corliss (1985) energy levels and provided a list 
of \ion{Mn}{ii} spectral lines and transition probabilities taken mostly from 
experimental sources. The \ion{Mn}{ii} atomic data of the NIST database
(Kramida et al. 2014)
are the result of the Kramida \& Sansonetti (2013) work. In the database
there are numerous unclassified lines of \ion{Mn}{ii}.
They are  concentrated in the 2380-2700\,\AA\ interval 
which was observed for HD\,175640 at about 120000 resolution. 
The new \ion{Mn}{ii} energy levels have permitted us to identify more 
than  50\% of the lines unclassified in the NIST database.

\section {Observations and data reduction}
  
HD\,175640 (HR\,7143) is one of the targets included in the ``Hot Stars'' program 
(GO-13346), which is part of the ``Advanced Spectral Library (ASTRAL)'' 
Project (Ayres 2014). The star was observed in  Cycle 21 of the Space 
Telescope Imaging Spectrograph (STIS). The spectrum covers the range 
1150-3045\,\AA. The nominal resolving power R ranges from
about 30000 to 120000; the S/N ratio is larger than 100. 
We used the final spectrum that resulted from the calibration and the merging of the individual
spectra observed in the different wavelength intervals, as performed by the ASTRAL Science 
Team (Carpenter et al. 2014).

For this work we analyzed the whole region from 1250 to 3045\,\AA.
We used the IRAF tool ``continuum'' to normalize the observed spectrum  to the continuum level.
When we compared observed and computed spectra we tentatively
fixed  different resolving powers for the different spectral intervals roughly corresponding
to the different resolutions of the different observed regions (Ayres 2010).
 In particular, we adopted R equal to
50000, 40000, 25000, 30000, and 120000 for the ranges 1250-1340\,\AA, 1340-1690\,\AA, 1690-2200\,\AA, 2200-2332\,\AA, 
and 2332-3045\,\AA, respectively.  
The observed spectrum was shifted to the Ritz wavelengths, as derived from the
energy levels, using mostly \ion{Fe}{ii} lines
as reference lines, owing to their accurate wavelengths derived from the Nave \& Johansson (2013)
energy levels. 
The shift to superimpose the observed spectrum to the synthetic spectrum is $+$25\,km\,s$^{-1}$ in
the 1250-2000\,\AA\ region. After conversion  of the wavelength scale of the 
2000-3040\,\AA\ interval from vacuum to air, the shift is $+$24.5\,km\,s$^{-1}$.
We estimate an  uncertainty in the velocity shift on the
order of 0.5\,km\,sec$^{-1}$, corresponding to an uncertainty 
in the wavelength scale of 0.2\,m\AA,
0.35\,m\AA, and 0.5\,m\AA\ at 1250\,\AA, 2000\,\AA, and 3000\,\AA, respectively. 

\begin{table}
\begin{center}
\caption{Abundances $\log$(N$_{elem}$/N$_{tot}$) for HD\,175640 [12000K,3.95]
adopted to compute the ultraviolet synthetic spectrum.
Solar abundances are from Asplund et al. (2009). }
\begin{tabular}{lrlrccccccccccccccc}
\hline\noalign{\smallskip}
\multicolumn{1}{c}{Elem.}&
\multicolumn{2}{c}{HD175640}&
\multicolumn{1}{c}{Sun}&
\\
\multicolumn{1}{c}{}&
\multicolumn{1}{c}{UV}&
\multicolumn{1}{c}{Visible}&
\multicolumn{1}{c}{}&
\\
\hline\noalign{\smallskip}
\ion{He}{i}  &    &  $-$1.73  & $-$1.11\\
\ion{Be}{ii} &    &  $-$10.64 & $-$10.99\\
\ion{B}{ii}   & $-$8.75  &          & $-$9.34\\
\ion{C}{i}   &$<$ $-$4.00 & $-$4.11$\pm$0.23 & $-$3.61 \\
\ion{C}{ii}   & $-$4.00  &  $-$4.05$\pm$0.16 & $-$3.61\\
\ion{N}{i}  & $-$5.95&     $\le$ $-$5.78     & $-$4.21\\
\ion{O}{i}  & $-$3.28&     $-$3.18$\pm$0.11  & $-$3.35\\
\ion{Ne}{i} &            &  $-$4.35          & $-$4.11\\
\ion{Na}{i} &            &  $-$5.47          & $-$5.80\\
\ion{Mg}{i} & $-$4.69    &  $-$4.64 $\pm$ 0.06 &$-$4.44\\
\ion{Mg}{ii} & $-$4.69   &  $-$4.71 $\pm$ 0.07 &$-$4.44\\
\ion{Al}{i}  &           &  $<$ $-$7.50      &$-$5.59\\   
\ion{Al}{ii} & $-$7.00   &                   & $-$5.59\\
\ion{Al}{iii}& $-$7.00   &                   & $-$5.59\\
\ion{Si}{i}   & $-$4.80 &                 & $-$4.53 \\
\ion{Si}{ii}  & $-$4.80    & $-$4.72 $\pm$0.08 & $-$4.53\\
\ion{Si}{iii} & $-$4.80    & $-$4.58 $\pm$0.04 & $-$4.53\\
\ion{Si}{iv}  & $-$4.80    &                   & $-$4.53\\
\ion{P}{i} & $<$ $-$6.28   &                   & $-$6.63\\
\ion{P}{ii}& $-$6.28       & $-$6.28$\pm$0.08  & $-$6.63\\
\ion{P}{iii} & $-$6.28     &                   & $-$6.63\\
\ion{S}{i}   & $-$5.30     &                   & $-$4.92\\
\ion{S}{ii}  & $-$5.30     & $-$5.12$\pm$0.03  & $-$4.92\\
\ion{Cl}{i} & $-$7.50      &                   & $-$6.54 \\
\ion{Ca}{i} &              & $-$5.26           & $-$5.70\\
\ion{Ca}{ii} & $-$5.54     & $-$5.67 $\pm$0.25 & $-$5.70\\
\ion{Sc}{ii} & $-$9.08     & $-$9.08 $\pm$0.15 & $-$8.89\\ 
\ion{Sc}{iii} & $-$9.08    &                   & $-$8.89\\
\ion{Ti}{ii}  &$-$5.67     & $-$5.67$\pm$0.11  & $-$7.09\\
\ion{Ti}{iii} & $-$5.67    &                   & $-$7.09\\
\ion{V}{ii}   & $-$9.94    & $\le$ $-$9.04     & $-$8.11\\
\ion{Cr}{i}   &            & $-$5.22 $\pm$0.09 & $-$6.40\\
\ion{Cr}{ii}  & $-$5.36    & $-$5.41 $\pm$0.07 & $-$6.40\\
\ion{Cr}{iii} & $-$5.36    &                   & $-$6.40\\
\ion{Mn}{i}   & $-$4.20    & $-$4.20$\pm$0.08  & $-$6.61\\
\ion{Mn}{ii}  & $-$4.20    & $-$4.25 $\pm$ 0.04& $-$6.61\\
\ion{Mn}{iii} & $-$4.20:   &                & $-$6.61\\
\ion{Fe}{i}   & $-$4.83    & $-$4.78$\pm$0.08  & $-$4.54\\
\ion{Fe}{ii}   & $-$4.83    & $-$4.84$\pm$0.13 & $-$4.54\\
\ion{Fe}{iii} & $-$4.83    &                   & $-$4.54\\
\ion{Co}{ii} & $-$9.00     & $-$8.08:          & $-$7.05\\
\ion{Ni}{ii} & $-$6.09     & $-$6.09$\pm$0.16  & $-$5.82\\  
\ion{Ni}{iii}& $-$6.09     &                   & $-$5.82\\
\ion{Cu}{i}  &             & $-$6.52           & $-$7.85\\
\ion{Cu}{ii} & $-$6.50     &                   & $-$7.85\\
\ion{Zn}{ii} & $-$8.70     &                   & $-$7.48\\
\ion{Ga}{i}  & $-$5.43 &                       & $-$9.00\\
\ion{Ga}{ii} & $-$5.43 &  $-$5.43$\pm$0.04     & $-$9.00\\
\ion{Ga}{iii} &$<$ $-$5.43 ? &                 & $-$9.00\\
\ion{Ge}{ii} & $-$10.1 &                      & $-$8.39\\
\ion{As}{ii} & $-$7.50 &                      & $-$9.74\\
\ion{Br}{ii} &         &  $-$7.12$\pm$0.04     & $-$9.50\\
\ion{Sr}{ii} &         &  $-$8.41              & $-$9.17\\
\ion{Y}{ii}  & $-$6.66 & $-$6.66$\pm$0.20      & $-$9.83\\
\ion{Y}{iii} & $>$ $-$6.66&                    & $-$9.83\\
\hline\noalign{\smallskip}
\end{tabular}
\end{center}
\end{table}

\setcounter{table}{0}

\begin{table}
\begin{center}
\caption{cont. }
\begin{tabular}{lrlrccccccccccccccc}
\hline\noalign{\smallskip}
\multicolumn{1}{c}{elem}&
\multicolumn{2}{c}{HD175640}&
\multicolumn{1}{c}{Sun}&
\\
\multicolumn{1}{c}{}&
\multicolumn{1}{c}{UV}&
\multicolumn{1}{c}{Visible}&
\multicolumn{1}{c}{}&
\\
\hline\noalign{\smallskip}
\ion{Zr}{ii}  &       & $-$8.67$\pm$0.17       & $-$9.46\\
\ion{Zr}{iii} & $-$8.50  &                     & $-$9.46\\
\ion{Rh}{ii}  & $-$8.50 & $-$8.50              & $-$11.13\\
\ion{Pd}{i}   &         &$-$6.41 $\pm$0.30     & $-$10.47\\ 
\ion{Pd}{ii}  & $-$6.30 &                      & $-$10.47\\
\ion{Pd}{iii}  & $-$6.30&                      & $-$10.47 \\
\ion{Ag}{ii}  & $-$8.50 &                      & $-$11.10\\
\ion{Cd}{ii}  & $-$9.53 &                      & $-$10.33\\
\ion{Xe}{ii}  &         &$-$5.96 $\pm$0.20 & $-$9.80 \\
\ion{Ba}{ii}  &         & $-$9.27          & $-$9.86\\
\ion{Pr}{iii} &         & $-$9.62         & $-$11.32  \\
\ion{Nd}{iii} &          & $-$9.57 $\pm$0.08 & $-$10.62\\
\ion{Yb}{ii}  & $-$8.10  & $-$8.10 $\pm$0.19 & $-$11.20\\
\ion{Yb}{iii}  & $>$ $-$8.10 & $-$7.31$\pm$0.01 & $-$11.20\\
\ion{Os}{ii}   &         & $-$10.55            & $-$10.64\\
\ion{Ir}{ii}  &  $-$11.15 & $-$10.66:          & $-$10.66\\ 
\ion{Pt} &  $-$10.42 &  $-$7.63      & $-$10.42\\
\ion{Au}{ii} & $-$7.51 & $-$7.51$\pm$0.06  & $-$11.12\\
\ion{Au}{iii}& $-$7.51 &                   & $-$11.12 \\
\ion{Hg}{i}  & $-$6.60 &  $-$6.19 $\pm$0.18 & $-$10.87\\
\ion{Hg}{ii} & $-$6.60 &  $-$6.53 $\pm$0.23 & $-$10.87\\
\ion{Hg}{iii} & $-$6.60&                    & $-$10.87 \\
\hline\noalign{\smallskip}
\end{tabular}
\end{center}
\end{table}

\section{The star HD\,175640}

The stellar parameters and the abundances of HD\,175640  were 
determined
by Castelli \& Hubrig (2004) in a previous analysis of the optical
spectrum of the star.   
An ATLAS12 (Kurucz 2005) model with parameters \teff=12000\,K, \logg=3.95, and
microturbulent velocity $\xi$=0\,km\,sec$^{-1}$ was computed  for the
individual stellar abundances. They were obtained from the equivalent widths
of selected lines and are listed in the third column of Table\,1. 
 The SYNTHE code (Kurucz 2005) was used to compute
a synthetic spectrum for the 3050 $-$ 10000\,\AA\ region which was compared
with the observed spectrum. The ATLAS12 model and the given set of abundances 
were adopted.  A rotational velocity of
 v{\it\,sini} $=$ 2.5\,km\,sec$^{-1}$ was derived from the comparison of
the observed and computed spectra.  The adopted zero microturbulent 
velocity was based both on 
the Adelman (1994) conclusions that most HgMn stars have little or no
microturbulent velocity and on the consistent abundances we obtained from 
the equivalent widths of 55 weak and strong \ion{Fe}{ii} lines 
measured in the optical region (Castelli \& Hubrig 2004).

In this paper we compared the observed HST-STIS  spectrum
with a synthetic spectrum computed for the 1250-3045\,\AA\ interval.
We used the ATLAS12 model atmosphere adopted in Castelli\& Hubrig (2004)
and the abundances  listed in the second column of Table\,1. We started
with the abundances adopted to compute the optical spectrum 
and then we modified some of them in order to better fit the
ultraviolet spectrum.
The abundances of \ion{B}, \ion{N}, \ion{Al},
\ion{Cl}, \ion{V}, \ion{Zn}, \ion{Ge}, \ion{As}, \ion{Ag}, and \ion{Cd} 
were derived only in this paper. 
 For those elements that do not show
lines in the HST-STIS spectrum we adopted the abundances derived from the optical
region if they did not produce inconsistent results, as is the case 
of \ion{Pt}.
For the ions \ion{C}{i}, \ion{P}{i}, \ion{Ga}{iii}, \ion{Y}{iii},
and \ion{Yb}{iii} the ionization equilibria are not satisfied. We did not use
their corresponding abundance to compute the synthetic spectrum, but that
of the ions in the other ionization stages.
In addition, the abundance derived from \ion{Mn}{i} and \ion{Mn}{ii}
gives incorrect profiles for a large number of \ion{Mn}{iii} lines, 
that are either not observed or are computed as too strong  (Figure\,1).
These lines are mostly due to transitions having either an
even level with  energy larger than 172000\,cm$^{-1}$ or  
an odd level with  energy  larger than 130000\,cm$^{-1}$.
We compared 
 the Kurucz $\log\,gf$ values that we used  with those from
Uylings \& Raassen (1997) for the lines in common.
Because the differences 
are not larger than 0.2\,dex (while fitting the observed lines 
requires lowering the $\log\,gf$ values by 1.0 up to 2.0\,dex or more),
we could argue that some physical mechanism like the vertical abundance
stratification  weakens
the high-excitation lines of \ion{Mn}{iii} in HD\,175640. On the other hand,  the
low-excitation lines, in particular those of multiplets
10, 11, 12, 13, 14, 15, 16, 17, and 18 (Moore 1950), which lie in
the 1900-2400\,\AA\ region, are reasonably well predicted by the adopted
abundance.

\begin{figure}
\centering
\resizebox{5.00in}{!}{\rotatebox{90}{\includegraphics[0,200][525,800]
{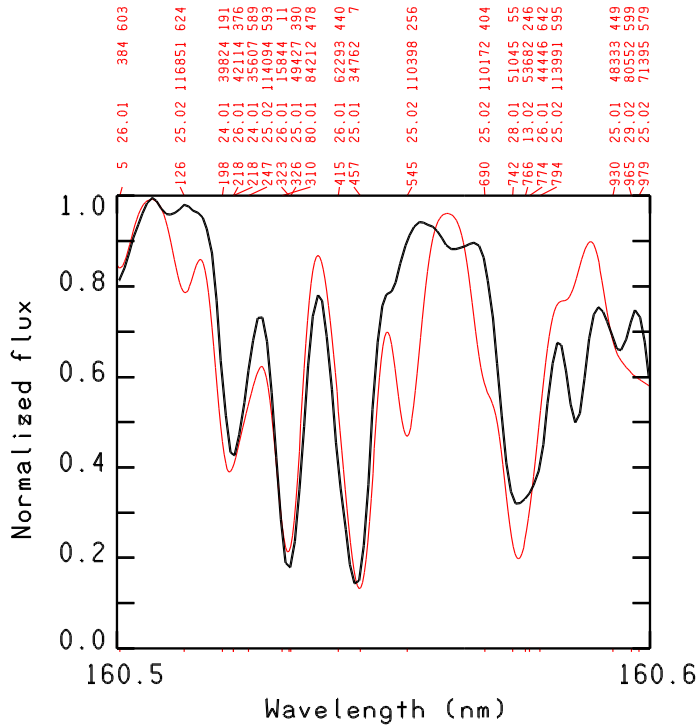}}}
\caption{Example of computed spectrum (red line) displaying \ion{Mn}{iii} lines
not observed in the stellar spectrum (black line). The lines are 
those at 1605.126\,\AA, 1605.545\,\AA, and 1605.690\,\AA. Their upper level
is an even level with energy 179152.090\,cm$^{-1}$, 172682.640\,cm$^{-1}$,
and 172451.230\,cm$^{-1}$, respectively. The labels for the lines include the last
three digits of the wavelength, the code (atomic number) for the element and ion,
the lower energy level in cm$^{-1}$, and the residual intensity at line
 center in per mil.}  
\end{figure}

To compute the synthetic spectrum  we used the Kurucz database
which is formed by the Kurucz (1988; 1993) line lists and by new and 
revised data (Kurucz 2011).
The Kurucz line lists include the hyperfine structure for several elements.
In particular, for \ion{Mn}{ii}, the magnetic dipole and electric
quadrupole coupling constants A and B measured by Holt et al. (1999) 
for 59 levels were
used to compute the hyperfine structure of all the lines with
coupling constants available for both levels.
The constants and the hyperfine lines are listed on Kurucz's website\footnote{
http://kurucz.harvard.edu/atoms/2501/hyper250155.pos or hyper250155.all}. 
The two files quoted in the footnote differ in that experimental $\log\,gf$ values,
when available, are used in the second file instead of the Kurucz computed
values which are stored in the first file.

For this work  we added lines of  \ion{Ga}{ii}, \ion{Ga}{iii}, 
\ion{Rh}{ii}, \ion{Pd}{ii}, \ion{Ag}{ii}, \ion{Y}{iii}, \ion{Zr}{iii}, \ion{Yb}{iii},
\ion{Pt}{ii}, \ion{Pt}{iii}, \ion{Au}{ii}, \ion{Au}{iii}, \ion{Hg}{ii},
\ion{Hg}{iii}. The presence of these elements in the STIS spectrum of 
HD\,175640 was inferred from the previous analysis of the optical region 
(Castelli \& Hubrig 2004). We also used  predicted lines and intensities for
the STIS region using the ``predicted stellar'' option of the VALD database
(Ryabchikova et al. 1997) with stellar parameters for HD\,175640 from
Castelli \& Hubrig (2004). Additionally, we used results from
wavelength coincidence statistic (WCS, Cowley \& Hensberge 1981),
using measurements of STIS wavelengths for $\lambda$ $>$ 2000\,\AA. These
wavelengths are available online\footnote{http://dept.astro.lsa.umich.edu/$\sim$cowley/HR7143air.html}.
More information on the heavy elements identified in HD\,175640 is given in 
the Appendix.

In addition to the large overabundance of \ion{Mn} ([$+$2.4]), 
\ion{Ti} ([$+$1.4]), \ion{Cr} ([$+$1.0]), and
the mild underabundance of \ion{Fe} ([$-$0.3]),  mild overabundances of 
\ion{B}, \ion{Na}, and \ion{P},  a general underabundance of the
light elements (He, C, N, Mg, Al, Si, S, Cl), and general overabundance of 
several heavy elements (Cu, Ga, As, Br, Sr, Y, Zr, Rh, Pd, Ag, Cd, Xe, Yb, Au, and Hg)
were measured. This peculiar chemical composition gives rise to a very rich 
line spectrum in the ultraviolet.

\section{New \ion{Mn}{ii} energy levels}

After adding to the Kurucz line list the missing atomic species 
relevant for the star and available in the literature, and after
comparing the observed spectrum with the synthetic spectrum computed with
the abundances given in Table\,1, we concentrated on the
\ion{Mn}{ii} lines. Line data for this ion were recently updated and critically
evaluated by Kramida \& Sansonetti (2013). They list 3969 \ion{Mn}{ii} lines
from 933.0682\,\AA\ to 9907.26\,\AA\ from which they derived
277 even energy levels and 377 odd energy levels.
 All the lines 
are classified as transitions between known energy levels, except for
188 lines observed in the laboratory spectra but  not classified.
Most of these lines lie in the 2380-2700\,\AA\ region.
Numerous unidentified lines can be observed in this region 
in HD\,175640. The wavelengths of several of them are the unclassified
\ion{Mn}{ii} lines.

To determine new energy levels we used the same method adopted 
by Castelli \& Kurucz (2010) when they
derived new \ion{Fe}{ii} energy levels from the optical UVES spectrum
of HR\,6000. 
Predicted energy levels and $\log\,gf$ values for \ion{Mn}{ii} were
computed by Kurucz with his version of the Cowan (1981) code  (Kurucz 2011).
The calculation included 50 even configurations d$^{6}$, d$^{5}$4d$-$12d,
d$^{4}$4s4d$-$4s9d, d$^{5}$4s$-$12s, d$^{4}$4s$^{2}$, d$^{4}$4s5s$-$4s9s,
d$^{5}$5g$-$9g, d$^{4}$4s5g$-$4s9g, d$^{5}$7i$-$9i, d$^{4}$4s7i$-$4s9i,
d$^{4}$4s9l, and d$^{4}$4p$^{2}$ with 19686 eigenvalues that were fitted
with the least-squares method to 277 known levels.  The 41 odd configurations included
d$^{5}$4p$-$10p, d$^{4}$4s4p$-$4s10p, d$^{3}$s$^{2}$4p,
d$^{5}$4f$-$10f, d$^{4}$4s4f$-$4s10f, d$^{5}$6h$-$9h,  d$^{4}$4s6h$-$4s9h,
d$^{5}$8k$-$9k, and d$^{4}$4s8k$-$4s9k with 19820 eigenvalues which were least-squares fitted 
to 377 known levels. The calculations were done in intermediate coupling with all 
configuration interactions included, with scaled Hartree-Fock starting
guesses, and with Hartree-Fock transition integrals.
A total of 5146779 lines were saved from the transition array of which 41882
lines are between known levels and have good wavelengths.

To derive new energy levels for \ion{Mn}{ii}, we considered predicted wavelengths 
due to transitions between a measured and a predicted energy level.
A predicted line is usually not shifted more than $\pm$10\,\AA\ from
the corresponding unknown observed line. However, owing to the large number 
of predicted lines, of unidentified lines, and blends in the ultraviolet
spectrum of HD\,175640,
it is not a straightforward task to find the correct correspondence between 
a predicted line and an unidentified line observed in the spectrum.

\begin{table}[htp]
\begin{flushleft}
\caption{New energy levels  of \ion{Mn}{ii} from the STIS-HST spectrum.}  
\begin{tabular}{rrcllllllllll}
\noalign{\smallskip}\hline
\multicolumn{2}{c}{Designation}&
\multicolumn{1}{c}{J}&
\multicolumn{1}{c}{Energy}
\\
&&&\multicolumn{1}{c}{cm$^{-1}$}&
\\
\noalign{\smallskip}\hline

3d$^{4}$($^{5}$D)4s4p($^{3}$P)& $^{7}$F$^{o}$ & 3.0 & 82303.415 \\
3d$^{4}$($^{5}$D)4s4p($^{3}$P)& $^{7}$F$^{o}$ & 4.0 & 82564.493 \\
\\
3d$^{5}$(b$^{2}$D)4p&$^{3}$F$^{o}$ & 2.0& 100320.586\\
                    &              & 4.0& 100725.176\\
\\
3d$^{5}$(b$^{2}$D)4p&$^{3}$P$^{o}$ & 1.0& 101807.469\\
\\
3d$^{6}$($^{4}$G)4d &$^{5}$F& 1.0& 106351.582 \\ 
                    &       & 2.0& 106348.454 \\
                    &       & 3.0& 106340.485 \\
                    &       & 4.0& 106324.758 \\
                    &       & 5.0& 106298.960 \\
\\
3d$^{5}$($^{4}$G)4d &$^{5}$G& 2.0 & 106408.011\\
                    &       & 3.0 & 106408.348\\
                    &       & 4.0 & 106406.642\\ 
                    &       & 5.0 & 106400.766\\
                    &       & 6.0 & 106387.310\\
\\
3d$^{5}$($^{4}$G)4d &$^{3}$D& 3.0 & 107192.929\\
\\
3d$^{5}$($^{4}$G)4d &$^{3}$I& 5.0 & 107155.877\\
                    &       & 6.0 & 107145 038\\ 
                    &       & 7.0 & 107114.165\\
\\
3d$^{5}$($^{4}$G)4d & $^{3}$G& 3.0& 107999.208\\
                    &        & 4.0& 108007.750\\
                    &        & 5.0& 108006.293\\
\\
3d$^{5}$($^{4}$G)4d & $^{3}$F& 2.0& 108420.629\\
                    &        & 3.0& 108425.105\\
                    &        & 4.0& 108420.479\\
\\
3d$^{6}$($^{4}$P)4d & $^{5}$P& 1.0& 108617.023\\
                    &        & 3.0& 108399.148\\
\\
3d$^{6}$($^{4}$P)4d &$^{5}$F & 1.0& 108511.206 \\
                    &        & 2.0& 108521.377\\
                    &        & 3.0& 108561.169\\
                    &        & 4.0& 108604.283\\
                    &        & 5.0& 108661.385\\
\\
3d$^{5}$($^{4}$G)4d &$^{3}$H & 4.0& 108907.450\\
                    &        & 5.0& 108906.437\\
                    &        & 6.0& 108896.070\\
\\
3d$^{6}$($^{4}$P)4d &$^{3}$D & 1.0& 109709.489\\
                    &        & 2.0& 109572.155\\
                    &        & 3.0& 109370.762\\
\\
3d$^{5}$($^{4}$P)4d &$^{3}$F & 2.0& 110664.620\\
                    &        & 3.0& 110611.014\\
\\
3d$^{5}$($^{4}$P)4d &$^{3}$P & 2.0& 110998.198\\
\\
3d$^{5}$($^{4}$D)4d &$^{5}$G & 2.0& 111744.523\\
                    &        & 3.0& 111753.977\\
                    &        & 4.0& 111761.546\\
                    &        & 5.0& 111761.008\\
                    &        & 6.0& 111741.583\\
\hline
\noalign{\smallskip}
\end{tabular}
\end{flushleft}
\end{table}

\setcounter{table}{1}

\begin{table}[htp]
\begin{flushleft}
\caption{Cont.}  
\begin{tabular}{rrcllllllllll}
\noalign{\smallskip}\hline
\multicolumn{2}{c}{Designation}&
\multicolumn{1}{c}{J}&
\multicolumn{1}{c}{Energy}
\\
&&&\multicolumn{1}{c}{cm$^{-1}$}&
\\
\noalign{\smallskip}\hline
 3d$^{5}$($^{4}$D)4d &$^{5}$P &3.0 & 111831.748\\
\\ 
3d$^{5}$($^{4}$D)4d &$^{5}$S & 2.0& 111991.133\\
\\
3d$^{5}$($^{4}$D)4d &$^{5}$F & 2.0& 112141.748\\
                    &        & 3.0& 112106.537\\
                    &        & 4.0& 112046.403\\
                    &        & 5.0& 111943.809\\
\\                   
3d$^{5}$($^{4}$D)4d &$^{5}$D & 4.0& 113199.572\\
\\
3d$^{5}$($^{2}$I)4d& $^{3}$I & 5.0& 118544.936\\
                   &         & 6.0& 118574.338\\
                   &         & 7.0& 118585.339\\
\\
3d$^{5}$($^{2}$I)4d& $^{1}$K & 7.0& 119152.797\\
\\
3d$^{5}$($^{2}$I)4d& $^{1}$G & 4.0& 119937.245\\
\\
\hline
\noalign{\smallskip}
\end{tabular}
\end{flushleft}
\end{table}

The computed line list was sorted into tables of all the lines with 
$\log\,gf$ $\ge$$-$3.0 connected to every predicted level. 
At first, we extracted the lines with $\log\,gf$ $\ge$$-$1.0 so that only 
the strongest lines were examined. When at least two strong predicted lines
originate from the same predicted level, we
searched in the spectrum for an unidentified line with wavelength close to the
first of the two predicted lines. From the observed wavelength and the known 
energy level involved in the transition we derived a possible value for 
the unknown energy level. We used this energy value to derive the 
corresponding wavelength of the second predicted line. 
If there is  an unidentified line at this position in the observed spectrum
we checked the energy value on other predicted lines connected to the
examined predicted level. If the test is positive for a sample of lines 
(usually from 3 to 5), we assign  the checked energy to the unknown level. 
Otherwise, we selected some other observed unidentified line in the 
spectrum and we repeated the procedure. We proceed in this way  until we 
find that value for the unknown energy that produces lines which are all 
observed in the spectrum,  but not identified.

Whenever one or more new levels was found, the whole semiempirical 
calculation was repeated to produce improved predicted wavelengths 
and $\log\,gf$ values.

Using the above procedure we identified 58 new energy levels from the HST-STIS 
spectrum of HD\,175640; five of them have odd 
parity, the others have even parity.  
We added 15 more new odd parity energy levels that we obtained
from the UVES spectrum of HD\,175640.
The new energy levels are listed in Table\,2 and Table\,3.

\begin{table}
\begin{flushleft}
\caption{New energy levels  of \ion{Mn}{ii} from the optical region (UVES spectrum).}  
\begin{tabular}{rrcllllllllll}
\noalign{\smallskip}\hline
\multicolumn{2}{c}{Designation}&
\multicolumn{1}{c}{J}&
\multicolumn{1}{c}{Energy}
\\
&& &\multicolumn{1}{c}{cm$^{-1}$}&
\\
\noalign{\smallskip}\hline
3d$^{5}$($^{6}$S)4f&$^{7}$F$^{o}$ & 1.0& 98423.93\\
                    &             & 0.0& 98424.00\\
\\
3d$^{5}$($^{6}$S)6f&$^{7}$F$^{o}$ & 2.0& 113840.6\\
                   &              & 1.0& 113840.7\\
                   &              & 0.0& 113840.8\\   
\\
3d$^{5}$($^{6}$S)8f&$^{7}$F$^{o}$ & 6.0& 119197.79\\
                   &              & 5.0& 119197.79\\
                   &              & 4.0& 119197.79\\
                   &              & 3.0& 119197.79\\
                   &              & 2.0& 119197.79\\
                   &              & 1.0& 119197.79\\
                   &              & 0.0& 119197.79\\
\\
3d$^{5}$($^{4}$G)4f&$^{5}$H$^{o}$ & 7.0& 125218.640\\
                   &              & 9.0& 125243.338\\
\\
3d$^{5}$($^{4}$G)4f&$^{5}$I$^{o}$ & 6.0& 125257.344\\ 
\\
\hline
\noalign{\smallskip}
\end{tabular}
\end{flushleft}
\end{table}

The search for new energy levels in the optical region was rather unsuccessful.
The energies  98423.93\,cm$^{-1}$ and 98424.00\,cm$^{-1}$ of the
two levels belonging to the term
3d$^{5}$($^{6}$S)4f\,$^{7}$F$^{o}$,
were obtained from the strong lines of the triplets 
at 5294\,\AA\ and 5295\,\AA, which are part of the multiplet  
3d$^{5}$($^{6}$S)4d\,e$^{7}$D$^{o}$ $-$ 3d$^{5}$($^{6}$S)4f\,$^{7}$F$^{o}$.
All the transitions of this multiplet give rise to five strong \ion{Mn}{ii}
lines observed at 5294, 5295, 5297, 5299, and 5302\,\AA.
Because the lines computed with the energies listed by 
Kramida \& Sansonetti (2013) are more or less displaced  from the observed 
lines, we modified the energy of the upper level in order to better 
fit the observed line position. Table\,4 compares  the energies  of
the 3d$^{5}$($^{6}$S)4f\,$^{7}$F$^{o}$ term adopted in 
this paper with those listed in the NIST database. 
The corresponding wavelengths of the multiplet are also listed.
Figure\,2 shows the five \ion{Mn}{ii} lines computed with
wavelengths from this paper listed in the first column of Table\,4
and those  from  Kramida\&Sansonetti (2013) listed in the 
last column of Table\,4.

\begin{figure*}
\centering
\resizebox{5.00in}{!}{\rotatebox{90}{\includegraphics[0,100][600,700]
{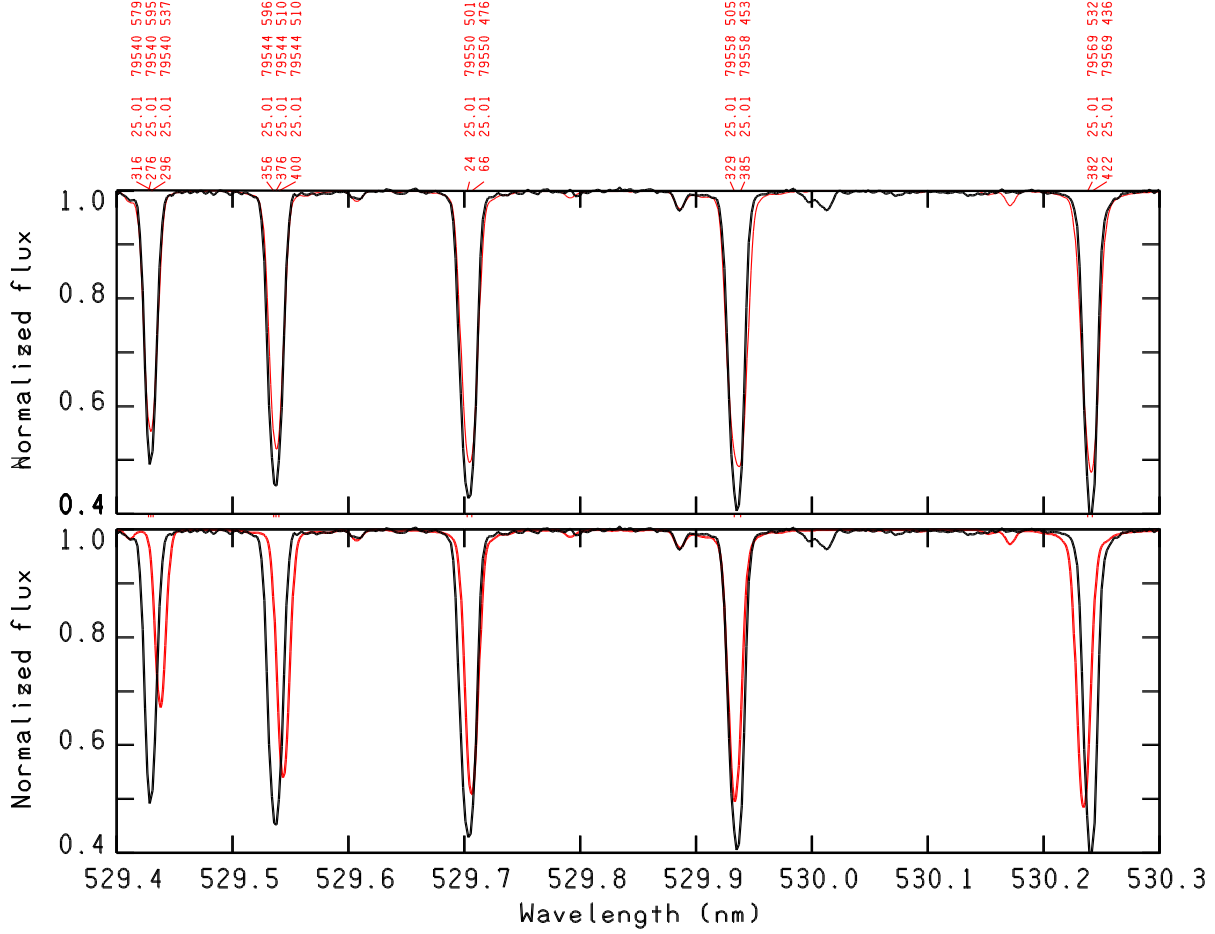}}}
\caption{Improved energy levels of \ion{Mn}{ii}.
Five lines from multiplet 
3d$^{5}$($^{6}$S)4d\,e$^{7}$D $-$ 3d$^{5}$($^{6}$S)4f\,e$^{7}$F$^{o}$ are 
computed using the wavelengths and the energies  from
Kramida \& Sansonetti (2013) (bottom panel), and the current values
(upper panel). The new and old wavelengths and energies are listed in
Table\,4. The labels are the same as in Fig.\,1.} 
\end{figure*}

The energies 113840.6\,cm$^{-1}$, 113840.7\,cm$^{-1}$, and 113840.8\,cm$^{-1}$ 
of the three levels with J=2, 1, and 0 respectively, belonging to the 
odd parity term
3d$^{5}$($^{6}$S)6f\,$^{7}$F$^{o}$ were simply extrapolated from the 
levels with
higher J quantum number. The transitions from these levels contribute to two
weak and blended lines observed at 7166.4 and 7167.4\,\AA.

The energy 119197.79\,cm$^{-1}$ was assumed for all the levels 
with J from 0 to 7 of the term 3d$^{5}$($^{6}$S)8f\,$^{7}$F$^{o}$.
 They give
rise to two weak lines observed at 5180.271\,\AA\ and 5181.649\,\AA. The first
one is blended with \ion{Fe}{ii} 5180.312\,\AA.

Finally, the observed lines arising from the three last energy levels 
of Table\,3 are listed in the Table A.1 available online.
In addition to the corrections of the odd parity levels shown in Table\,4, we modified
two other odd parity energy levels from Kramida \& Sansonetti (2013) on  the basis 
of all the transitions observed in the spectrum related to them.
They are: 81802.746\,cm$^{-1}$ instead of
81803.31\,cm$^{-1}$ (3d$^{5}$\,($^{2}$I)4p\,z$^{1}$I$^{o}$, J=6) and 
94230.94\,cm$^{-1}$ instead of 94231.20\,cm$^{-1}$ 
(3d$^{5}$\,($^{2}$S)4p\,z$^{3}$P$^{o}$, J=2).   

\begin{table}
\caption[ ]{The multiplet 3d$^{5}$($^{6}$S)4d\,e$^{7}$D $-$ 3d$^{5}$($^{6}$S)4f\,e$^{7}$F$^{o}$. The energies and wavelengths from this paper are compared
with those from Kramida\& Sansonetti (2013). We note that with the new levels,
all 15 of the possible lines of the $^{7}$D$-$$^{7}$F multiplet are known.} 
\font\grande=cmr7
\grande
\begin{flushleft}
\begin{tabular}{lllllrrcllllll}
\hline\noalign{\smallskip}
\multicolumn{1}{c}{$\lambda$(CKC)$^{a}$}&
\multicolumn{1}{c}{El(KS)$^{b}$}&
\multicolumn{1}{c}{Eu(CKC)$^{a}$} &
\multicolumn{1}{c}{Jl,Ju}&
\multicolumn{1}{c}{Eu(KS)$^{b}$}&
\multicolumn{1}{l}{$\lambda$(KS)$^{b}$}
\\
\multicolumn{1}{c}{$\AA$}&
\multicolumn{1}{c}{cm$^{-1}$}&
\multicolumn{1}{c}{cm$^{-1}$}&
\multicolumn{1}{c}{}&
\multicolumn{1}{c}{cm$^{-1}$}&
\multicolumn{1}{c}{$\AA$}\\
\hline\noalign{\smallskip}
    5294.276 & 79540.93 & 98424.00 &1,0 & $---$   & $---$\\
    5294.296 & 79540.93 & 98423.93 &1,1 & $---$   & $---$\\
    5294.318 & 79540.93 & 98423.858&1,2& 98423.63 & 5294.379\\
\\
    5295.356 & 79544.71 & 98423.93 &2,1&\\
    5295.376 & 79544.71 & 98423.858&2,2& 98423.63 & 5295.439\\
    5295.400 & 79544.71 & 98423.773&2,3& 98423.63 & 5295.439\\
\\
    5297.003 & 79550.50 & 98423.858 &3,2& 98423.63 & 5297.063\\
    5297.024 & 79550.50 & 98423.773 &3,3& 98423.63 & 5297.063\\
    5297.068 & 79550.50 & 98423.624 &3,4& 98423.60 & 5297.072\\
\\
    5299.288& 79558.56  & 98423.773 &4,3& 98423.63 & 5299.327\\ 
    5299.329& 79558.56  & 98423.624 &4,4& 98423.60 & 5299.336\\
    5299.387& 79558.56  & 98423.424 &4,5& 98423.59 & 5299.338\\
\\
    5302.327& 79569.22  & 98423.624 &5,4& 98423.60 & 5302.331\\
    5302.382& 79569.22  & 98423.424 &5,5& 98423.59 & 5302.333\\
    5302.421& 79569.22  & 98423.281 &5,6& 98423.56 & 5302.342\\
\hline
\noalign{\smallskip}
\end{tabular}
\end{flushleft}
$^{a}$ CKC:this paper;\\
$^{b}$ KS: Kramida \& Sansonetti (2013)\\
\end{table}

\section{The new \ion{Mn}{ii} lines}

The new \ion{Mn}{ii} lines due to transitions from the new \ion{Mn}{ii}
energy levels listed in Table\,2 and Table\,3 are given 
in  Table\,A.1, available online.  The new \ion{Mn}{ii}
lines are mostly concentrated in the 2380-2700\,\AA\ interval.
The upper energy levels (Cols. 1$-$4) were derived as described in Sect.\,4;
the lower energy levels (Cols. 5$-$8) were taken from
Kramida \& Sansonetti (2013); the wavelength given in Col.\,9 is the
Ritz wavelength in air for $\lambda$ $\ge$ 2000\,\AA, in vacuum for
$\lambda$$<$2000\,\AA. The $\log\,gf$ values (Col.\,10) were computed
by Kurucz with the semi-empirical method, the  wavelengths given in 
Col.\,11 are the observed laboratory wavelengths listed by 
Kramida \& Sansonetti (2013). In addition to the 109 newly classified lines,
we give a different assignment for  24 other lines of the 
Kramida \& Sansonetti (2013) tabulation. This implies different energy levels
for these 24 lines. The last column of Table\,A.1 indicates whether the
new line is listed in the NIST database without any classification (unassigned),
listed with a different classification (DIFFERENT assign.), or  not
listed at all.

Figure\,3 shows the \ion{Mn}{ii} spectrum in the regions 2383-2390\,\AA\
interval computed before and after the determination of
both new energy levels and new \ion{Mn}{ii} lines. 
Figure\,4 compares
the observed spectrum of HD\,175640 with the synthetic spectrum computed
both with and without the new \ion{Mn}{ii} lines.  
The improvement
of the comparison between the observed and computed spectra 
due to the addition of the new \ion{Mn}{ii} lines is evident.

\begin{figure*}
\centering
\resizebox{5.00in}{!}{\rotatebox{90}{\includegraphics[0,100][500,700]
{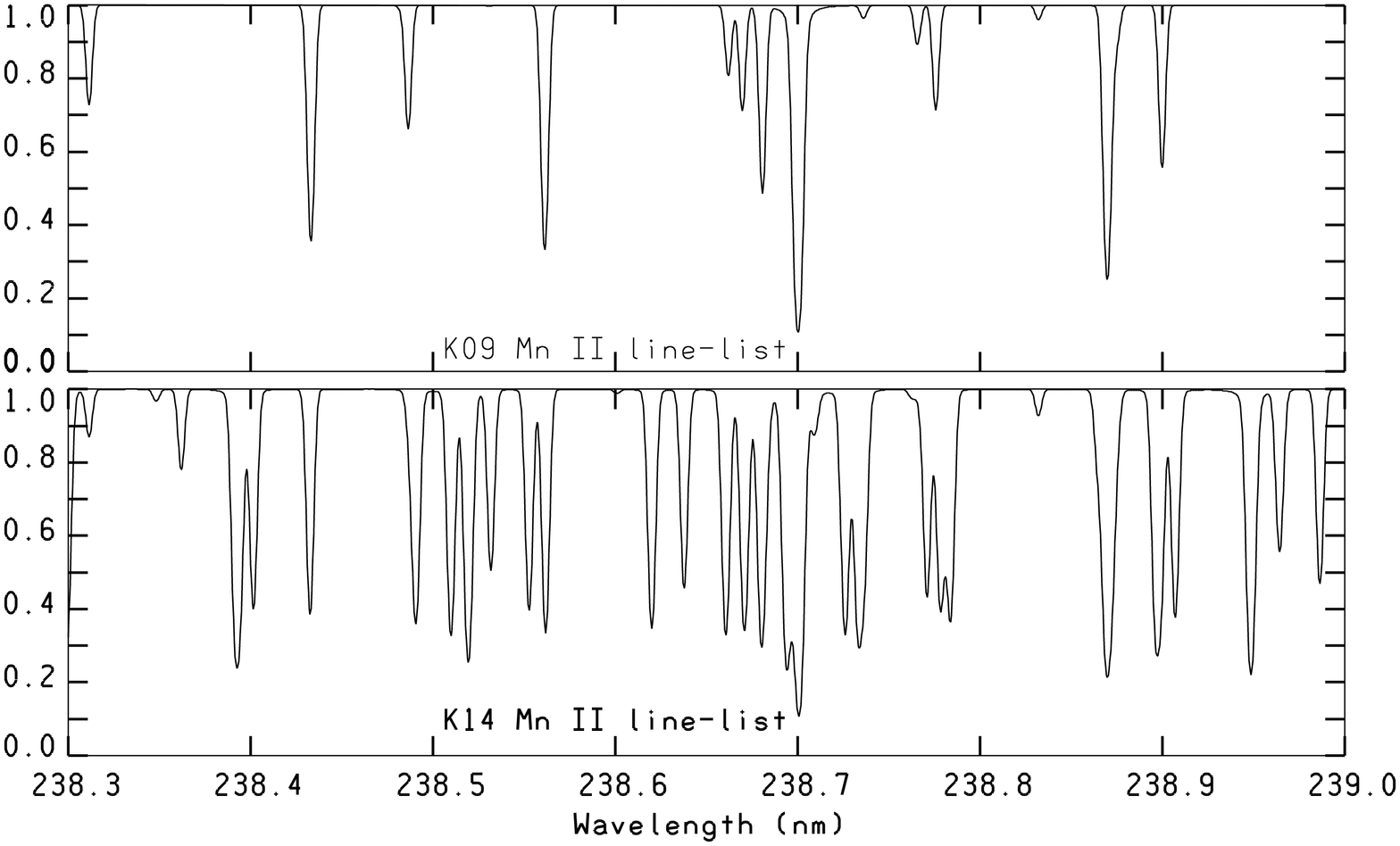}}}
\caption{Upper panel: \ion{Mn}{ii} synthetic spectrum for the
parameters of HD\,175640 (\teff=12000\,K, \logg=3.95, v{\it sini}=2.5\,km$^{-1}$, 
[Mn/H]]=+2.4) computed with the \ion{Mn}{ii} line list computed by Kurucz
before this work (2009 line list); 
lower panel: the same, but with the new \ion{Mn}{ii} lines added in the line list.
}
\end{figure*}

\begin{figure*}
\centering
\resizebox{5.00in}{!}{\rotatebox{90}{\includegraphics[0,100][600,700]
{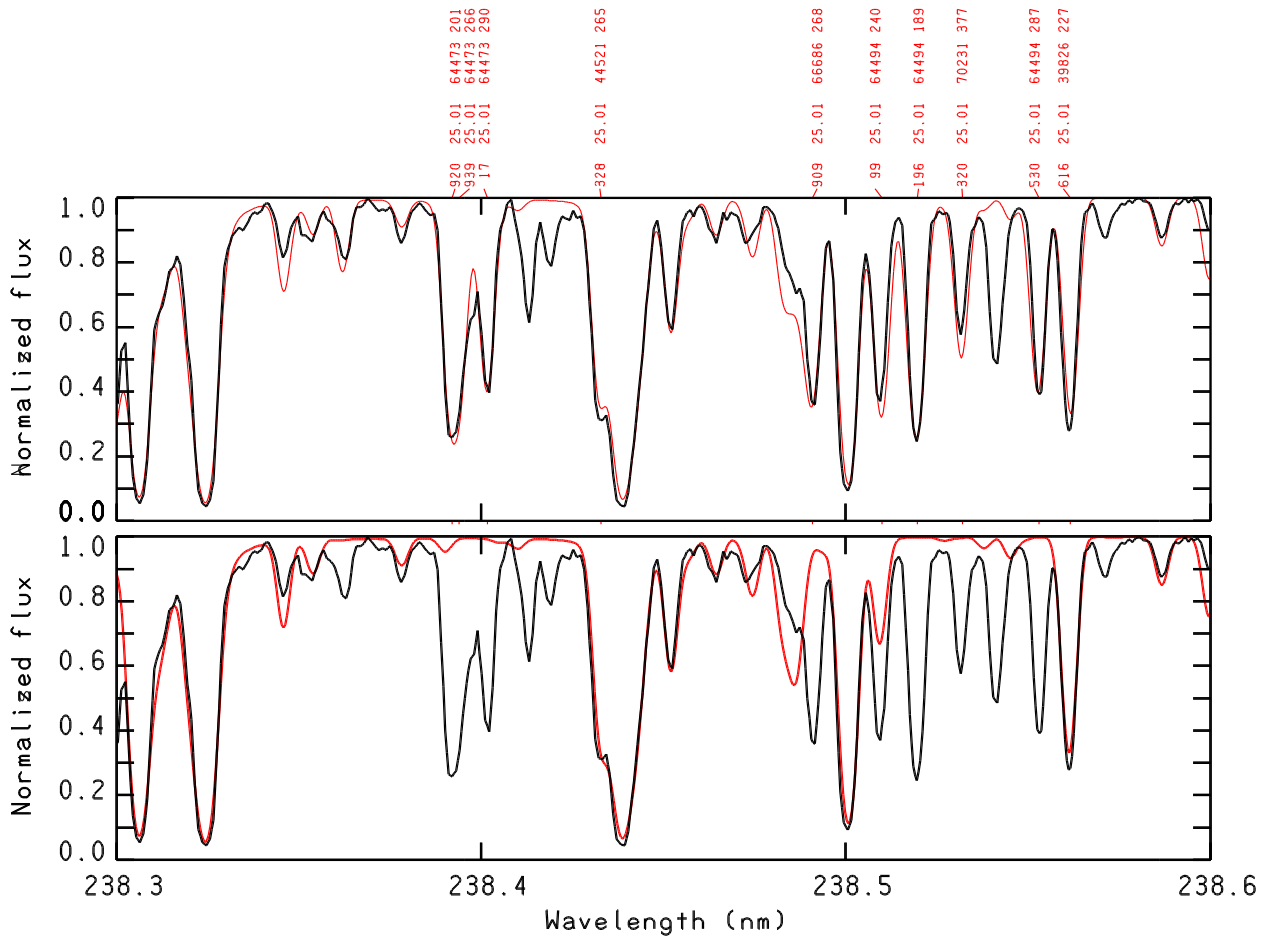}}}
\caption{Upper panel: Comparison of the STIS spectrum of HD\,175640 (black line)
with a synthetic spectrum (red line) computed with a line list including the new \ion{Mn}{ii} lines;
lower panel: the same, but the synthetic spectrum is computed with a line list not
including the lines given in Table\,A.1 available online. Only the \ion{Mn}{ii} lines are indicated  with red
labels in the plot. The meaning of the labels is the same as in Fig.\,1.}
\end{figure*}

\begin{table*}
\caption[ ]{Abundances from \ion{Mn}{ii} lines with experimental and semi-empirical $\log\,gf$ values
that differ less than $\pm$0.06\,dex.} 
\font\grande=cmr7
\grande
\begin{flushleft}
\begin{tabular}{rrrllrrcllllll}
\hline\noalign{\smallskip}
\multicolumn{1}{c}{wave}&
\multicolumn{1}{c}{$\log\,gf$}&
\multicolumn{1}{c}{E(low)}&
\multicolumn{1}{c}{J(low)}&
\multicolumn{1}{c}{Term}&
\multicolumn{1}{c}{E(up)}&
\multicolumn{1}{c}{J(up)}&
\multicolumn{1}{c}{Term}&
\multicolumn{1}{c}{hfs}&
\multicolumn{1}{c}{$\log\,gf$}&
\multicolumn{1}{c}{Acc.}&
\multicolumn{1}{c}{$\Delta\log\,gf$}&
\multicolumn{1}{l}{Abund}
\\
\multicolumn{1}{c}{$\AA$}&
\multicolumn{1}{c}{K14}&
\multicolumn{1}{c}{cm$^{-1}$}&
\multicolumn{1}{c}{ }&
\multicolumn{1}{c}{ }&
\multicolumn{1}{c}{cm$^{-1}$}&
\multicolumn{1}{c}{}&
\multicolumn{1}{c}{ }&
\multicolumn{1}{c}{}&
\multicolumn{1}{c}{KSG}&
\multicolumn{1}{c}{}&
\multicolumn{1}{c}{K14-KSG}&
\multicolumn{1}{l}{}\\

\hline\noalign{\smallskip}

1678.643& $-$2.681 &  9472.993 & 2.0& ($^{6}$S)4s\,a\,$^{5}$S & 69044.910 & 2.0 & ($^{4}$P)4p\,z$^{3}$P& no & $-$2.638&D$^{+}$& $-$0.043& $-$4.20\\
2535.659& $-$0.478 & 27583.590 & 4.0& ($^{4}$G)4s\,a\,$^{5}$G & 67009.217 & 3.0 & ($^{4}$P)4p\,z$^{5}$D& no & $-$0.502&B$^{+}$& $+$0.024& $-$3.80\\
2535.977& $-$1.024 & 27588.534 & 3.0& ($^{4}$G)4s\, a\,$^{5}$G & 67009.217 & 3.0 & ($^{4}$P)4p\,z$^{5}$D& no & $-$1.045&B$^{+}$& $+$0.021& $-$3.80\\
2543.457& $-$0.119 & 27589.360 & 2.0& ($^{}$4G)4s\, a\,$^{5}$G & 66894.130 & 1.0 & ($^{4}$G)4p\,z$^{5}$F& no & $-$0.138&B& $+$0.019& $-$4.20\\
2557.543& $-$0.310 & 27588.534 & 3.0& ($^{4}$G)4s\, a\,$^{5}$G & 66676.833 & 2.0 & ($^{4}$P)4p\,z$^{5}$D& no & $-$0.346&C$^{+}$& $+$0.036& $-$4.10\\
2716.793& $-$0.664 & 29889.534 & 3.0& ($^{4}$P)4s\, a\,$^{5}$P & 66686.739 & 3.0 & ($^{4}$G)4p\,z$^{5}$F& yes & $-$0.646&B$^{+}$& $-$0.018& $-$3.80\\
2762.558& $-$1.422 & 32857.270 & 3.0& ($^{4}$D)4s\, b\,$^{5}$D & 69044.910 & 2.0 & ($^{4}$P)4p\,z$^{3}$P& no & $-$1.471&D$^{+}$& $+$0.049& $-$4.20\\
2933.785& $-$1.421 & 32818.440 & 0.0& ($^{4}$D)4s\, b\,$^{5}$D & 66894.130 & 1.0 & ($^{4}$G)4p\,z$^{5}$F& yes & $-$1.458&B& $+$0.037& $-$4.20\\
2935.362& $-$2.047 & 32836.740 & 1.0& ($^{4}$D)4s\, b\,$^{5}$D & 66894.130 & 1.0 & ($^{4}$G)4p\,z$^{5}$F& no  & $-$2.002&C$^{+}$& $-$0.045& $-$3.85\\
2949.204& $+$0.313 &  9472.993 & 2.0& ($^{6}$S)4s\, a\,$^{5}$S & 43370.537 & 3.0 & ($^{6}$S)4p\,z$^{5}$P& no & $+$0.253$^{a}$&A$^{+}$& $+$0.060& $-$4.20\\
2955.139& $-$0.972 & 32857.270 & 3.0& ($^{4}$D)4s\, b\,$^{5}$D & 66686.739 & 3.0 & ($^{4}$G)4p\,z$^{5}$F& no & $-$0.987&B$^{+}$& $+$0.015& $-$3.80\\
2989.730& $-$1.934 & 33248.660 & 4.0& ($^{4}$G)4s\, a\,$^{3}$G & 66686.739 & 3.0 & ($^{4}$G)4p\,z$^{5}$F& no & $-$1.924&D$^{+}$& $-$0.010& $-$3.90\\
3050.657& $-$0.159 & 36274.620 & 2.0& ($^{4}$P)4s\, b\,$^{3}$P & 69044.910 & 2.0 & ($^{4}$P)4p\,z$^{3}$P& no & $-$0.206&B$^{+}$& $+$0.047& $-$4.15\\
3146.121& $-$2.546 & 34910.770 & 4.0&  d6\, b\,$^{3}$3G & 66686.739 & 3.0 & ($^{4}$G)4p\,z$^{5}$F& no & $-$2.523&D& $-$0.023& $-$4.25\\
3204.878& $-$1.115 & 37851.490 & 3.0&  d6\, a\,$^{3}$3D & 69044.910 & 2.0 & ($^{4}$P)4p\,z$^{3}$P& no & $-$1.097&B$^{+}$& $-$0.018& $-$4.15\\ 
4206.368& $-$1.584 & 43528.661 & 5.0& ($^{4}$F)4s\,a\,$^{5}$F & 67295.446 & 4.0 & ($^{4}$P)4p\,z$^{5}$D& yes& $-$1.553&C$^{+}$& $-$0.031& $-$4.05\\
4292.233& $-$1.581 & 43395.395 & 3.0& ($^{2}$D)4s\,c\,$^{3}$D & 66686.739 & 3.0 & ($^{4}$G)4p\,z$^{5}$F& yes& $-$1.544&D$^{+}$& $-$0.037& $-$4.20 \\
4343.983& $-$1.105 & 43528.661 & 5.0& ($^{4}$F)4s\,a\,$^{5}$F & 66542.539 & 5.0 & ($^{4}$G)4p\,z$^{5}$F& yes& $-$1.109&C$^{+}$& $+$0.004& $-$3.80\\
\hline
\noalign{\smallskip}
\end{tabular}
$^{a}$ This value is from Den Hartog et al. (2011)\\
\end{flushleft}
\end{table*}

\section{\ion{Mn}{ii} $\log\,gf$ values and  manganese abundance}

For \ion{Mn}{ii}, Kramida \&Sansonetti (2013) tabulate in the NIST database 
the 
experimental $\log\,gf$ values from  Den Hartog et al. (2011), 
Kling \& Griesmann (2000), and  Kling, Schnabel \& Griesmann (2001). For lines with upper levels 3d$^{5}$($^{6}$S)4p\,z$^{5}$P$^{o}$
and 3d$^{5}$($^{6}$S)4p\,z$^{7}$P$^{o}$, they adjusted the $\log\,gf$ values given in the last 
two papers according to the lifetime value 
recommended by Den Hartog et al. (2011) for these levels. Furthermore, they adopted 
the semi-empirical $\log\,gf$ values from Kurucz (1988) for several lines.

\begin{table*}
\caption[ ]{\ion{Mn}{ii} lines with experimental and semi-empirical $\log\,gf$ values
that differ more than $\pm$1.00\,dex.} 
\font\grande=cmr7
\grande
\begin{flushleft}
\begin{tabular}{rrlllrrcllllll}
\hline\noalign{\smallskip}
\multicolumn{1}{c}{wave}&
\multicolumn{1}{c}{$\log\,gf$}&
\multicolumn{1}{c}{E(low)}&
\multicolumn{1}{c}{J(low)}&
\multicolumn{1}{c}{Term}&
\multicolumn{1}{c}{E(up)}&
\multicolumn{1}{c}{J(up)}&
\multicolumn{1}{c}{Term}&
\multicolumn{1}{c}{hfs}&
\multicolumn{1}{c}{$\log\,gf$}&
\multicolumn{1}{c}{Acc.}&
\multicolumn{1}{c}{$\Delta\log\,gf$}&
\multicolumn{1}{l}{$\log\,gf$ for the}&
\\
\multicolumn{1}{c}{$\AA$}&
\multicolumn{1}{c}{K14}&
\multicolumn{1}{c}{cm$^{-1}$}&
\multicolumn{1}{c}{ }&
\multicolumn{1}{c}{ }&
\multicolumn{1}{c}{cm$^{-1}$}&
\multicolumn{1}{c}{}&
\multicolumn{1}{c}{ }&
\multicolumn{1}{c}{}&
\multicolumn{1}{c}{KSG}&
\multicolumn{1}{c}{}&
\multicolumn{1}{c}{K14-KSG}&
\multicolumn{1}{l}{$-$4.20\,dex abund}
\\
\hline\noalign{\smallskip}
2559.416& $-$0.350 & 27583.590 & 4.0& ($^{4}$G)4s\,a\,$^{5}$G & 66643.296 & 4.0 & ($^{4}$G)4p\,z$^{5}$F& no & $-$1.424&D& $+$1.074& $-$0.35\\
2717.524& $-$0.895 & 29889.534 & 3.0& ($^{4}$P)4s\, a\,$^{5}$P & 66676.833 & 2.0 & ($^{4}$P)4p\,z$^{5}$D& yes& $-$2.184&E& $+$1.289& $-$0.70\\
2719.736& $-$0.319 & 29919.444 & 2.0& ($^{4}$P)4s\,a\,$^{5}$P & 66676.833 & 2.0 & ($^{4}$P)4p\,z$^{5}$D& yes& $-$1.500&D& $+$1.181& $-$0.319&\\
2845.848& $-$2.732 & 31514.710 & 4.0&  d6\, a\,$^{3}$F & 66643.296 & 4.0 & ($^{4}$G)4p\,z$^{5}$F& no & $-$1.274&D$^{+}$& $-$1.458& $-$2.85\\
2956.005& $-$1.325 & 32857.270 & 3.0& ($^{4}$D)4s\,b\,$^{5}$D & 66676.833 & 2.0 & ($^{4}$P)4p\,z$^{5}$D& no & $-$3.168&D& $+$1.843& $-$0.75\\
2993.603& $-$2.421 & 33147.710 & 5.0& ($^{4}$G)4s\, a\,$^{3}$G & 66542.539 & 5.0 & ($^{4}$G)4p\,z$^{5}$F& no & $-$1.061&D$^{+}$& $-$1.360& $-$2.35&\\
4325.047& $-$2.355 & 43528.661 & 5.0& ($^{4}$F)4s\, a\,$^{5}$F & 66643.296 & 4.0 & ($^{4}$G)4p\,z$^{5}$F& yes& $-$1.115& C$^{+}$&$-$1.240& $-$2.55\\
4393.385& $-$2.646 & 44139.031 & 1.0& ($^{2}$D)4s\, c\,$^{3}$D & 66894.130 & 1.0 & ($^{4}$G)4p\,z$^{5}$F& yes& $-$1.458&D$^{+}$& $-$1.188& $-$2.70\\
\hline
\noalign{\smallskip}
\end{tabular}
\end{flushleft}
\end{table*}

 The comparison of the experimental $\log\,gf$ values (KGS) 
 with the last version of the semi-empirical
oscillator strengths from Kurucz (2014) (K14) is shown in  Fig\,5.
The average of the difference $\log\,gf$(K14)$-$$\log\,gf$(KGS) is
$+$0.068$\pm$0.397. The trend is small, but the dispersion around
the zero is rather large. However, on a total of 193 lines, 51 of them
have a difference in $\log\,gf$ less than 0.1\,dex.

\begin{figure*}
\centering
\resizebox{5.00in}{!}{\rotatebox{90}{\includegraphics[0,100][400,700]
{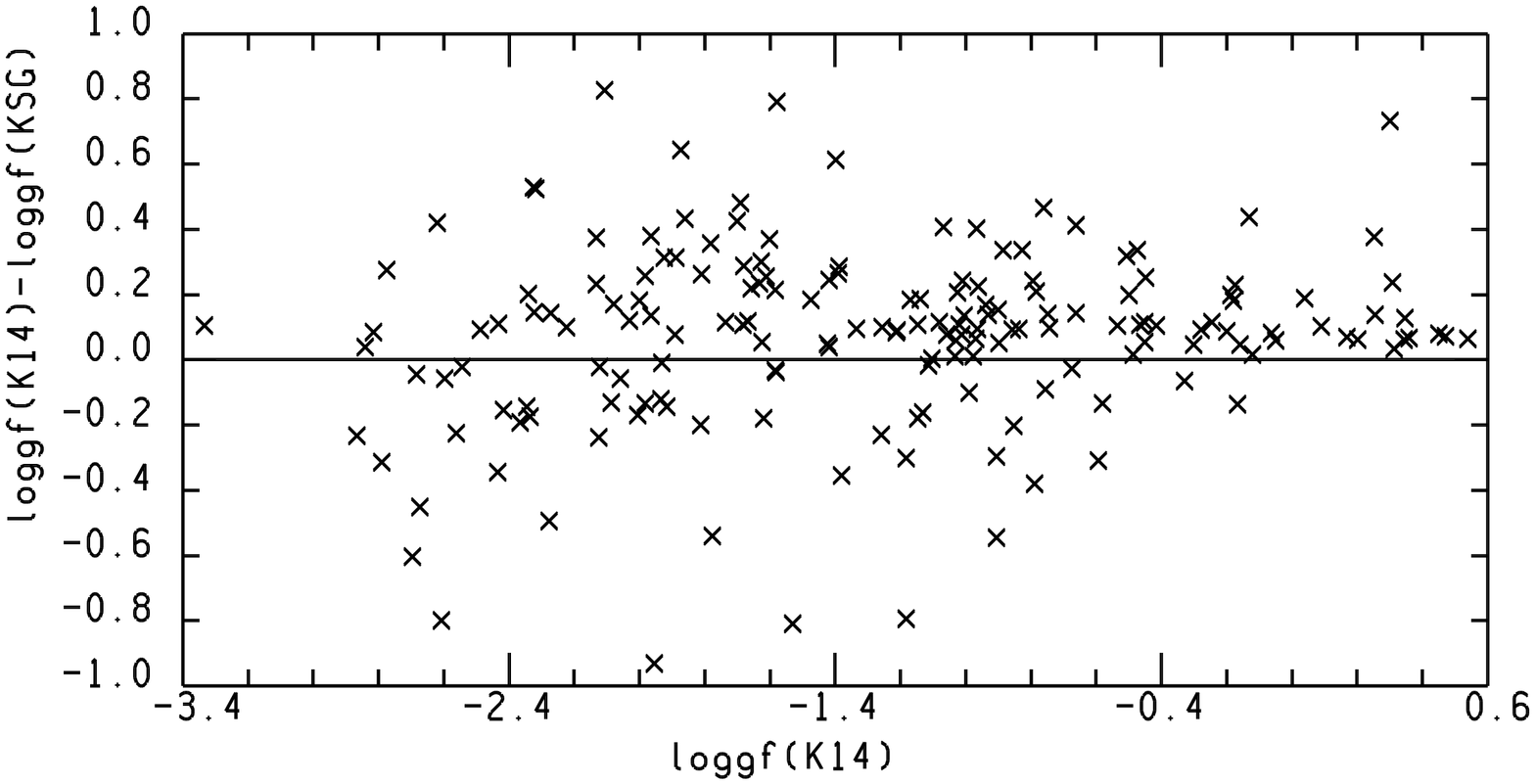}}}
\caption{Comparison of the experimental $\log\,gf$ values from Kling\& Griesmann (2000)
and Kling et al. (2001) with the semi-empirical computed values from Kurucz (2014).
There are still eight points outside the figure, four with $\log\,gf$$>$1.0 and four
with $\log\,gf$$<$$-$1.0. 
} 
\end{figure*}

In order to extract a set of lines well suited to deriving the manganese abundance,
we presumed that lines with small differences 
between experimental and semi-empirical  $\log\,gf$ values could be used
to this purpose.
Table\,5 lists 18 \ion{Mn}{ii} lines with $\log\,gf$ values 
from K14 and from either KGS or Den Hartog et al. (2011) differing less than 0.06\,dex.
The smallest difference (0.004\,dex) occurs for the line at 4343.983\,\AA. 
This line gives an abundance of $-$3.80\,dex.
We checked this abundance on the other lines listed in Table\,5.
Instead of deriving approximately the same abundance from the lines, we 
found a variety of values ranging from $-$3.8\,dex to $-$4.2\,dex, as 
is shown in the last column of Table\,5.
 We investigated whether the abundance differences are related to the  uncertainty 
associated with the experimental $\log\,gf$ values. According to
Kramida \& Sansonetti (2013) their accuracy  
is $\le$ 0.009\,dex for the score  A$^{+}$ and decreases up to  $\le$ 0.24\,dex for the score D. 
If we restrict our analysis to the lines in Table\,5 with the
scores B,  B$^{+}$ and A$^{+}$, corresponding
to uncertainties not greater than 0.04\,dex, we find that the discrepancies in the
abundances are not reduced. Figure\,6 shows  that the problem occurs because 
some strong \ion{Mn}{ii} lines have profiles that cannot be reproduced
by computations for any adopted abundance. For instance, while
the high abundance of $-$3.8\,dex improves the agreement between the   
observed and computed profiles of the strong line with no wings at 
2716.793\,\AA, it is too high for the broad line at 2949.204\,\AA, which
develops unobserved strong wings for abundances larger
than $-$4.20\,dex. We note that the experimental $\log\,gf$ value 
of this line is determined with an uncertainty $\le$0.009\,dex, so that it
cannot be the cause of the weaker than observed computed core for the
$-$4.20\,dex abundance, well suited to fitting the wings.
An increase in the microturbulent velocity from 0.0\,km\,sec$^{-1}$
to higher values does not solve the problem because the effect is the same
as that due to an increase in the abundance.
The difficulty in fitting the profile of several strong lines can be due
to missing hyperfine
structure (but this is not the case of the line at 2716.793\,\AA);
$\log\,gf$ uncertainties larger than those estimated;
the adopted model, which may have an incorrect stratification in the uppermost
layers; and some manganese vertical abundance stratification causing a 
manganese accumulation in the upper layers. 

The analysis of all the \ion{Mn}{ii} lines in HD\,175640 from 1250\,\AA\ to 10000\,\AA,
has shown that the most reliable manganese abundance is that derived
from the lines of multiplet 1 in the 3438-3500\,\AA\ spectral region.
The measured equivalent widths listed in Castelli \& Hubrig (2004) together with the
$\log\,gf$ values from Den Hartog et al. (2011) give an average abundance of
$-$4.17$\pm$0.03\,dex. This abundance reproduces very well both the wings and the
core of the lines at 3441.896, 3460.315, 3482.904, and
3488.676\,\AA. For the other lines of the multiplet the observed core is a little stronger
than the computed value.

There are eight points outside the borders of Fig.\,5, which were not plotted.
They correspond to the lines listed in Table\,6  for which the difference 
between the K14 and KSG $\log\,gf$ values is larger than $\pm$1.0\,dex.
The uncertainties of the experimental $\log\,gf$ values is high because it is
included between $\le$0.08\,dex (score C$^{+}$) and $>$0.24\,dex (score E).
We note that the source of the $\log\,gf$ values for the first three lines in Table\,6
was erroneously  indicated in the NIST database as c88 (Kurucz 1988) instead 
of T7259 (Kling et al. 2001).
The $\log\,gf$ value needed to reproduce the line profile when the abundance
of $-$4.20\,dex is adopted is given in the last column of Table\,6.
For all the lines, this $\log\,gf$ value is closer to that from K14
than to that  from KGS. This result suggests that, for some lines, the experimental $\log\,gf$ values
may be less reliable than the calculated values.

\begin{figure*}
\centering
\resizebox{5.00in}{!}{\rotatebox{90}{\includegraphics[0,160][540,660]
{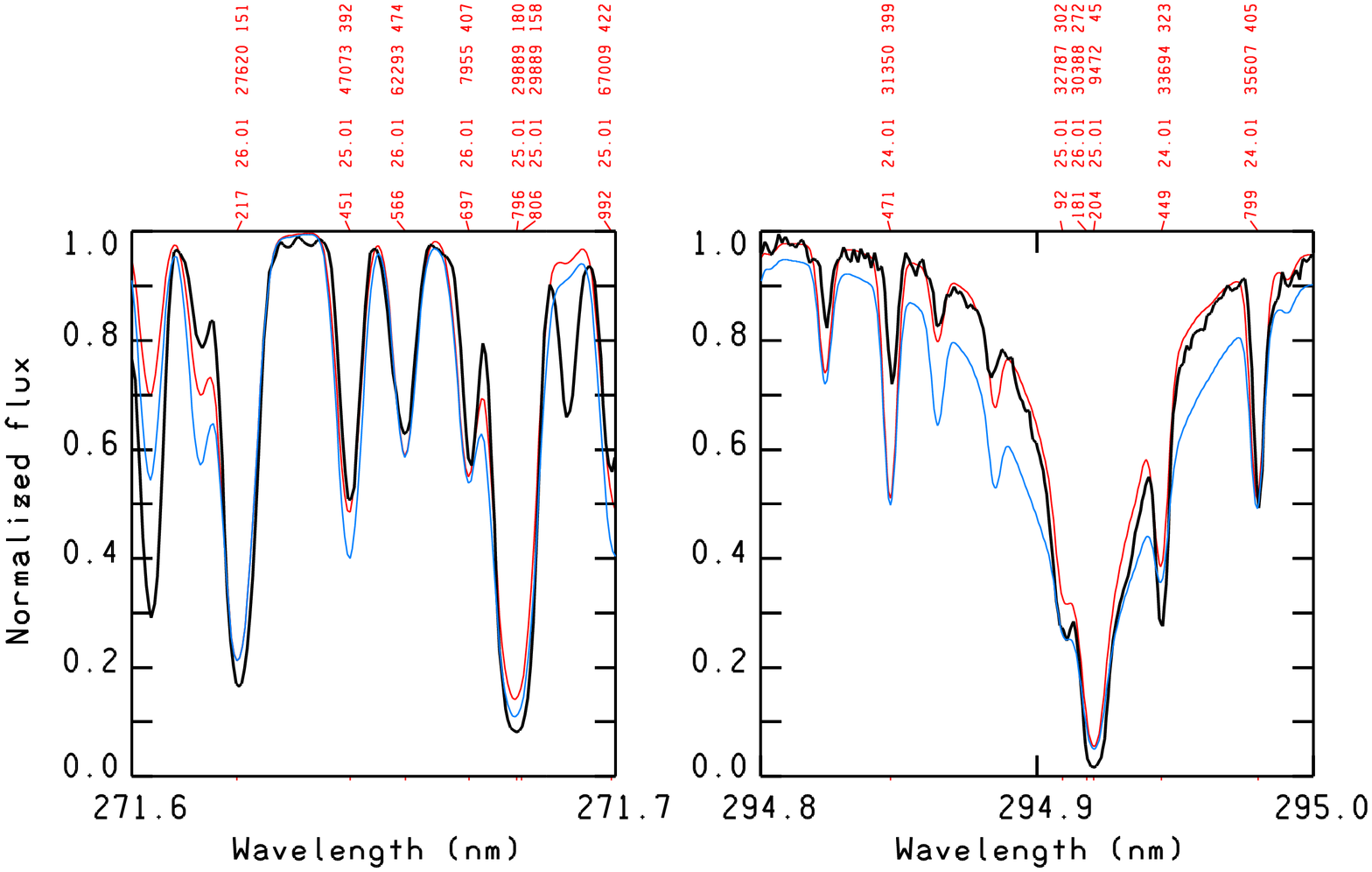}}}
\vskip -1cm
\caption{The comparison of the observed spectrum (black line) with synthetic
spectra computed for a manganese abundance of $-$4.2\,dex (red line) and
$-$3.8\,dex (blue line). The plot on the left shows how the computed core
of the line at 2716.793\,\AA\ increases with increasing abundance, although
it does not fit the observed spectrum even with $-$3.8\,dex. A still larger
abundance would develop unobserved wings. The line is computed with the hyperfine
structure included. The strongest hyperfine components are those at 2716.796\,\AA\ and
2716.806\,\AA. The plot on the right shows how the wings
of the strong \ion{Mn}{ii} line at 2949.204\,\AA\ are well fitted 
by the profile computed with the abundance of $-$4.2\,dex. The higher abundance 
gives rise to unobserved wings. The meaning of the labels is the same as in Fig.\,1.   
} 
\end{figure*}

\section{Comparison of the observed and computed spectra}

The plot of the comparison of the observed and computed spectra of HD\,175640
in the whole 1250-3040\,\AA\ region is available at the Castelli
website\footnote{http://wwwuser.oats.inaf.it/castelli/hd175640stis/tab1250-3040.html}.

When we compare the whole observed region from 1250\,\AA\ to 3046\,\AA\ 
with the synthetic spectrum  we are far from satisfied.
From 1250\,\AA\ to 1692\,\AA\ the agreement is poor, except for a few lines. 
In fact, in spite of the rather high resolving power  
(R $\sim$ 40000 $-$ 50000), the blends are so numerous and formed by so many
components that it is not easy to compute them correctly.
There are numerous missing lines, lines computed that are too strong, lines computed
that are too weak, and lines  probably affected by small  wavelength errors. 
Furthermore, for a few elements, the abundance is different for the different
ionization states. 
Between 1692\,\AA\ and 2332\,\AA\ the comparison is even worse because
all the above problems are increased by the rather low resolving power 
(R $\sim$ 25000$-$30000). 
Only in the 2332 $-$ 3040\,\AA\ region does the comparison show a rather good
agreement. This range is less crowded with lines
than the previous regions and was observed with a very high resolving power
($\sim$ 120000).

\subsection{Problematic \ion{Mn}{ii} lines}

If we restrict the discussion only to the \ion{Mn}{ii} lines, we see that
there are several lines with computed profiles without an observed counterpart.
For instance, all
the lines arising from the even parity level 43131.51\,cm$^{-1}$ (3d$^{6}$\,a$^{1}$D,
J=2) are  predicted as being strong,
but they are either not observed at all or have a very weak observed 
counterpart.  
Another remarkable line is that at 2703.213\,\AA, which is predicted as very 
strong with both the  Kurucz (2014) and  Kling et al. (2001) $\log\,gf$ values,
i.e. $-$0.484 and $-$0.808, respectively, but is not observed
at all in the spectrum. It would be too long to list here all the
\ion{Mn}{ii} lines computed as too strong compared to the observed ones.
In some cases, a possible explanation for the disagreement is the 
severe mixing affecting one of the two energy levels involved 
in the transition and the difficulty in computing it. Furthermore,
it is very likely that the computed stellar atmosphere is inadequate to explain
all the lines observed in the spectrum, in particular for manganese which
is very overabundant. In addition to the shortcoming affecting 
the core of the strong \ion{Mn}{ii} lines which, as discussed in Sect.\,6, 
is computed as too weak compared to the observed core,  there are 
several \ion{Mn}{ii} lines in the optical spectrum that seem to be 
weakened by some emission.
While the computed lines are strong  the observed lines are weak or even 
absent. All these lines are due to transitions between high-energy levels
like the blend at 
6446.188,  6446.247, 6446.330\,\AA, the lines at 6462.220,
6462.465, 6462.800, 6463.200, 6463.627\,\AA, those at 9903.853, 
9904.428, 9905.216, 9906.2, and 9907.2\,\AA, among others.
 
On the other hand, there are  a few strong lines observed in the spectrum
that are identified as \ion{Mn}{ii} in the NIST database, but which are 
either not predicted by the Kurucz (2014) $\log\,gf$ values
or are predicted as much too weak. Table\,A.2, available online,
lists the most significant lines.

\subsection{Absorption components of resonance lines}

A few strong resonance lines  are either double with a red component 
or are very broad with a redshifted line core. In this last case,
the profile is the blend  of the line predicted at the Ritz wavelength
with an unpredicted redshifted line.
In Table\,7 we list in Col.\,2 the Ritz wavelength and
in Cols.\,3 and 4 the wavelengths $\lambda_{obs}$
and $\lambda_{comp}$  of the two components
when they can be disentangled. In this case $\lambda_{obs}$ is the
same as the Ritz wavelength, otherwise $\lambda_{obs}$ is the wavelength of
the blend of the two components unless the line is so saturated that the core
lies below the zero level of the normalized flux scale and cannot be
measured. The velocity shift of the component
is given in Col.\,5. We estimate an error of about 2.5\,km\,sec$^{-1}$ related
to an uncertainty of 0.02\,\AA\ in the position of the component on the
wavelength scale. The average velocity shift is
8.29$\pm$0.56\,km\,s$^{-1}$.
We note that the \ion{Zn}{ii} line at 2062.001\,\AA\ is predicted but not observed,
 while
an unidentified line is present at 2062.060\,\AA. We assumed that it is
the red component of \ion{Zn}{ii} 2062.001\,\AA. However, it could be the
unpredicted line of some  other element as well.

Other similar peculiarities observed in HD\,175640 are the
bump and the broad weak component 
affecting the red wing of the \ion{Ca}{ii} profiles at 3933\,\AA\ (K-line) and
3968\,\AA\ (H-line), respectively. Furthermore, strong 
red components can be observed for the \ion{Na}{i} lines at 5890\,\AA\
and 5896\,\AA\ (Castelli \& Hubrig, 2004). 
We measured a velocity shift for the \ion{Na}{i} components equal 
to $+$7.4$\pm$0.7\,km\,sec$^{-1}$ and $+$7.7$\pm$0.7\,km\,sec$^{-1}$,
respectively. 

We cannot say whether the redshifted components are  of interstellar
or circumstellar origin. 

\begin{table}
\caption[ ]{Lines affected by a redshifted component.} 
\font\grande=cmr7
\grande
\begin{flushleft}
\begin{tabular}{lrlllllcllllll}
\hline\noalign{\smallskip}
\multicolumn{1}{c}{Elem}&
\multicolumn{1}{c}{$\lambda$(Ritz)}&
\multicolumn{1}{c}{$\lambda$(obs)}&
\multicolumn{1}{c}{$\lambda_{comp}$} &
\multicolumn{1}{c}{V$_{shift}$} &
\multicolumn{1}{c}{Notes}&
\\
\hline\noalign{\smallskip}
\multicolumn{1}{c}{}&
\multicolumn{1}{c}{($\AA$)}&
\multicolumn{1}{c}{($\AA$)}&
\multicolumn{1}{c}{($\AA$)}&
\multicolumn{1}{c}{km\,s$^{-1}$}&
\multicolumn{1}{c}{}&
\\
\hline\noalign{\smallskip}
\ion{Mg}{i} & 2852.126 &2852.126 &2852.20& 7.78& double\\
\ion{Mg}{ii}& 2795.528 &         &        &     & saturated\\
            & 2802.705 &         &        &     & saturated\\
\ion{Si}{ii}& 1808.013 &1808.03  &        &     & \\
\ion{S}{ii} & 1250.584 &1250.60  &        &     & \\
\ion{Mn}{ii}& 2576.104 &2576.104 & 2576.18& 8.84& double\\
            & 2593.721 &2593.721 & 2593.80& 9.13& double \\
            & 2605.680 &2605.68  & 2605.75& 8.05& double \\
\ion{Fe}{ii}& 2343.495 &2343.53  &        &     & \\
            & 2373.735 &2373.735 & 2373.80&8.21&double\\
            & 2382.037 &         &        &     & saturated\\
            & 2585.876 &2585.876 & 2585.94& 7.42&double\\
            & 2599.395 &         &        &     & saturated\\
\ion{Zn}{ii}& 2025.484 &2025.51  &        &     &  \\ 
            & 2062.001 & $--$    & 2062.06&8.57& \\
\hline
\noalign{\smallskip}
\end{tabular}
\end{flushleft}
\end{table}

\section{Conclusions}

Ultraviolet and optical stellar spectra of the HgMn slowly rotating star
HD\,175640, observed with both HST-STIS
and UVES instruments,  were used to extend and discuss the atomic data of
\ion{Mn}{ii} available in the NIST database (Kramida \& Sansonetti, 2013).
To this purpose,  \ion{Mn}{ii} lines arising both from levels observed in the
laboratory and from levels predicted with semi-empirical methods by Kurucz,
 were adopted. 
 
We assigned wavelengths, energy levels and $\log\,gf$ values to 
about 257 \ion{Mn}{ii} lines. 
Of them, 109 lines had already been identified as \ion{Mn}{ii} by 
Kramida \& Sansonetti (2013), but were unclassfied.  
For another 24 lines we assigned different energy levels from those in 
the NIST database. This implies different energy levels,
different $\log\,gf$ values, and therefore a different intensity for
the lines. 
The new \ion{Mn}{ii} line data improve the computation of the synthetic spectra
of B-type stars, although more work needs to be done on the atomic data,
especially in the ultraviolet. For instance, we note that in the NIST database
only wavelengths and intensities are listed for the \ion{Mn}{iii} lines
of the 1250-3040\,\AA\ region analyzed in the paper. No energy levels and no
$\log\,gf$ values are given for them. 

As by-product of this study we have extended to the ultraviolet region 
the abundance analysis performed by Castelli \& Hubrig (2004) in the 
optical region. Except for  cobalt, iridium, and platinum,  
we confirmed the optical abundances within the error limits of the equivalent
widths analysis made by Castelli\& Hubrig (2004).
Abundances for elements not observed in the 
visible, such as \ion{B}, \ion{N}, \ion{Al}, \ion{Cl}, \ion{V}, \ion{Zn}, 
\ion{Ge}, \ion{As}, \ion{Ag}, and \ion{Cd}  were obtained.
 The abundance pattern 
is similar to that of other HgMn stars which have underabundances of the
light elements and overabundances of some iron group elements and of
 some heavy elements. In the case of HD\,175640, the most
overabundant elements are \ion{Ti}, \ion{Cr}, \ion{Mn}, \ion{Ga}, 
\ion{As}, \ion{Br}, \ion{Y}, \ion{Zr}, \ion{Rh}, \ion{Pd}, \ion{Ag}, 
\ion{Xe}, \ion{Yb}, \ion{Au}, and \ion{Hg}.

\appendix

\section{Heavy elements  in HD\,175640}

We based the abundance analysis in the ultraviolet mostly on 
the lines listed by Castelli et al. 
(1985) and by Castelli \& Bonifacio (1990) in their study  
of HR\,6000 and $\iota$\,Her, respectively. The multiplet numbers given 
in the above  papers refer to the Ultraviolet Multiplet Table by Moore (1950).
In addition, we made a wide use of the NIST database (Kramida et al. 2014) 
to search for the most intense lines of a given element.
In this section we give some more details only about heavy elements 
from copper to mercury. \\
{\bf Copper (Cu) Z=29}: The abundance of $-$6.50\,dex was estimated 
from the \ion{Cu}{ii} lines at 1358.773, 1367.951, 1472.395, 2112.100, 
2135.981, 2192.268, 2242.618, and 2247.003\,\AA. All the lines 
are blended. The $\log\,gf$ values are data computed by  Kurucz in 2011 
and were taken from his database.\\   
{\bf Zinc (Zn) Z=30}: The zinc abundance was derived from the line of
\ion{Zn}{ii} at 2064.227\,\AA. The $\log\,gf$ value $+$0.070 is 
from the NIST database.
The \ion{Zn}{ii} line at 2062.001\,\AA\ originating from a lower level
with energy equal to 0.00\,cm$^{-1}$ is predicted, but not observed.
The $\log\,gf$ value $-$0.329 is from the Kurucz database. The other
line from the zero energy level at 2025.484\,\AA\ is computed too weak
and is blended with a strong unidentified line that we assumed to be
of interstellar or circumstellar origin. No \ion{Zn}{iii} lines were 
observed.\\
{\bf Gallium (Ga) Z=31}: For \ion{Ga}{i}, \ion{Ga}{ii}, and \ion{Ga}{iii}, 
wavelengths, energy levels, and $\log\,gf$ values were taken from the 
NIST database when available. 
For \ion{Ga}{i}, only the lines at 2874.23, 2943.64, and 2944.17\,\AA\ 
were well observed. 
For the \ion{Ga}{ii} lines with $\log\,gf$ values not available in the
NIST database, we used  $\log\,gf$ values from Nielsen et al. (2005)
(lines at at 1463.576, 1473.690, 1483.453, and 1483.903\,\AA).  
For the remaining lines $\log\,gf$ values were computed from the lifetime 
measuraments of  Ansbacher et al. (1985) or estimated on the basis
of laboratory intensities and excitation energies 
(Castelli \& Parthasarathy 1995). 
The  abundance $-$5.43\,dex ([+3.6]) derived from the optical region
adequately reproduces the unblended \ion{Ga}{ii} lines at 1473.690, 1483.903,
1504.334, 1536.276\,\AA, as well as the blended lines 
at 1514.505, 1535.312\,\AA.
The \ion{Ga}{ii} broad line at 1414.399\,\AA\ has too narrow wings, but 
the classical broadening parameters adopted here are probably not correct.
Vice versa, both \ion{Ga}{iii} profiles at 1495.045 and 1534.462\,\AA\ display
wings that are too broad. Only these two \ion{Ga}{iii} lines have $\log\,gf$
values available in the NIST database. For other lines we adopted 
estimated values.
No isotopic and hyperfine structure was considered in the computations.\\ 
{\bf Germanium (Ge) Z=32}: Several \ion{Ge}{ii} lines were observed.  From the
line at 1649.19\,\AA\ we derived an underabundance of 1.7\,dex relative to
the solar value. We adopted loggf= $-$0.28 from the NIST database\\  
{\bf Arsenic (As)  Z=33)}: The lines of \ion{As}{ii} at 1263.77, 1266.34, 1280.987,
and 1287.54\,\AA\ were observed in the spectrum. We adopted
$\log\,gf$ values from Warner\& Kirkpatrick (1969) for all them. 
The observed lines are adequately predicted for an overabundance of $-$7.50\,dex.\\ 
{\bf Yttrium  (Y) Z=39}: For the \ion{Y}{ii} lines, we used the 
Kurucz database, which includes hyperfine components for several lines.
All the lines are blended, except for the strong line at 2422.18\,\AA.
The abundance of $-$6.66\,dex [+3.2] derived from the optical spectrum is well
suited to reproducing the observed profile.
We added to the line list the \ion{Y}{iii} lines from Bi\'emont et al. (2011).
We modified the two wavelengths 2817.037\,\AA\ and 2946.01\,\AA\ in 
2817.027\,\AA\ and 2945.995\,\AA, in order
to match the observed spectrum. The \ion{Y}{iii} lines at 2367.228,
2414.643, 2817.027, and 2945.995\,\AA\ were observed. They are
strong lines, either unblended or marginally blended.
 The abundance from \ion{Y}{iii} is 
larger than the abundance from \ion{Y}{ii} by about 1.0\,dex.\\ 
{\bf Zirconium (Zr) Z=40}:  
Several weak \ion{Zr}{iii} lines were observed in the spectrum,
but no \ion{Zr}{ii} line. A few \ion{Zr}{iii} lines are unblended,
like that at 2102.283\,\AA.
The \ion{Zr}{iii} line data from the NIST database were adopted.
The \ion{Zr}{iii} abundance agrees, within the error limits, 
with the value we obtained from the \ion{Zr}{ii} optical lines.\\ 
{\bf Rhodium (Rh) Z=45}: Numerous weak \ion{Rh}{ii} lines were observed 
in the spectrum. The line at 1604.45\,\AA\ is unblended.
Oscillator strengths from B\"ackstr\"om et al. (2013), 
Quinet et al. (2012), and Corliss\& Bozman (1962) were used.
The wavelengths from the NIST database agree more closely with the stellar
wavelengths than the wavelengths adopted by B\"ackstr\"om et al. (2013).
The difference is on the order of 0.02\,\AA.
The abundance of $-$8.50\,dex, which was estimated from the UVES spectra on
the basis of estimated $\log\,gf$ values, reproduces the ultraviolet lines
in a satisfactory way. None of the \ion{Rh}{iii} lines listed in the NIST database was
identified.\\ 
{\bf Palladium (Pd) Z=46}: Numerous \ion{Pd}{ii} lines were identified. All  the lines from the
4d$^{8}$5s$-$4d$^{8}$5p transitions and the strongest lines from the
4d$^{8}$5p$-$4d$^{8}$6s transitions, together with their oscillator strengths, 
from Quinet (1996) were added in the line lists.
Additional lines were taken from  Lundberg et al. (1996) and from  the NIST database. 
Unblended lines are those at 2351.347, 2367.966, 2388.310, 2414.7303, 2426.867,
2433.102, 2446.713, 2457.257, 2472.502, 2486.256, 2488.914, 2505.729,
2565.505, 2569.544, 2635.93, and 2658.72\,\AA.
None of the \ion{Pd}{i} lines listed in the NIST database was identified.
For \ion{Pd}{iii}, only the line at 1782.55\,\AA, which according to the
NIST data is the one
with the largest intensity in the 1250-3040\,\AA\ range, 
perhaps contributes to a strong blend observed at 1782.6\,\AA.\\ 
{\bf Silver (Ag) Z=47}: The \ion{Ag}{ii} lines at 2246.412, 2248.749, 
2357.917, and
2411.345\,\AA\ were observed. The $\log\,gf$ values from 
the NIST database were used when available, otherwise we used 
the Corliss \& Botzman (1962) data, as we did for the line at 
2438.325\,\AA, which is computed as too strong.\\
{\bf Cadmium (Cd) Z=48}: The \ion{Cd}{ii} lines at 2144.393\,\AA\ and 2265.019\,\AA\ 
were observed and predicted by adopting a 0.8 overabundance over the
solar value. The atomic data are from the Kurucz database.
The $\log\,gf$ values are the same as in the NIST database.\\ 
{\bf Indium (In) Z=49}: The \ion{In}{ii} line at 1586.331\,\AA, which is  predicted
as rather strong for solar abundance  and $\log\,gf$ value
from the NIST database, is the main component
of a  complex blend formed by several other lines. We were unable to 
determine the indium abundance from the blend.\\ 
{\bf Barium (Ba) Z=56}: The two strongest \ion{Ba}{ii} lines at 2304.249\,\AA\
and 2335.267\,\AA\ are heavily blended, and so  they cannot be used
to confirm the abundance derived from the optical region. No other 
\ion{Ba}{ii} lines were observed.\\
{\bf Ytterbium (Yb) Z=70}: We adopted the abundance of $-$8.10\,dex ([+3.0])
derived from the UVES spectra.
We investigated only the \ion{Yb}{ii} lines with $\log\,gf$ available
in the NIST database. The lines at
2185.716, 2653.745, and 2750.478 can be observed as single weak lines.
Some other lines are weak components of blends.
We added to the line list  all the \ion{Yb}{iii} lines
from  Bi\'emont et al. (2001) 
with an upper energy level lower than 100000\,cm$^{-1}$.
There are numerous \ion{Yb}{iii} lines in the spectrum. Their profile
is very sharp and would require an abundance about 1\,dex larger than
that derived from the \ion{Yb}{ii} lines in order to agree with the 
computed profiles. The same behavior for \ion{Yb}{ii} and
\ion{Yb}{iii} abundances was observed in the visible
(Castelli \& Hubrig 2004).\\ 
{\bf Osmium (Os) Z=76}: No \ion{Os}{ii} lines were observed\\ 
{\bf Iridium (Ir) Z=77:} 
Except for the weak line observed
at 2245.750\,\AA, no other \ion{Ir}{ii} lines can be
conclusively identified in the spectrum. From this
line we derived an abundance of $-$11.15\,dex, i.e  
an underabundance of [-0.5]. The $\log\,gf$ values for \ion{Ir}{ii}
were taken from  the Kurucz database.
None of the \ion{Ir}{ii} lines listed in Ivarsson et
al. (2004) was observed in the spectrum, even assuming a
solar abundance.
The \ion{Ir}{ii} line at 3042.553\,\AA\
is a minor contributor of a blend with \ion{Ti}{ii}.
The blending was not considered when we analyzed 
the UVES spectrum, so that we derived a solar iridium
abundance (Castelli \& Hubrig 2004).\\
{\bf Platinum (Pt) Z=78}: For \ion{Pt}{ii} and \ion{Pt}{iii} we adopted the lines and the 
transition probabilities listed by Wyart et al. (1995)
and by Ryabtsev et al. (1993), respectively.
We did not find any clear evidence for a {Pt} overabundance.
Only the \ion{Pt}{ii} line at 1777.086\,\AA\ is predicted with solar abundance 
and observed. It is a blend with \ion{Fe}{ii} 1777.058\,\AA, which is computed 
as weaker than observed, but an overabundance of platinum 
gives rise to computed lines not observed in the spectrum.
We believe that the line at 4514.124\,\AA, 
identified as \ion{Pt}{ii} in the UVES spectrum (Castelli \& Hubrig 2004),
is actually due to some other element.\\ 
{\bf Gold (Au) Z=79}: Gold is  overabundant by 3.6\,dex. The abundance from the optical region
adequately reproduces the
numerous \ion{Au}{ii} lines observed in the ultraviolet. A few of them are unblended, 
good examples being the \ion{Au}{ii} lines at 1469.142,  1673.587, 1740.475,
1793.297 (blend), and 1800.579\,\AA. There are several predicted \ion{Au}{iii} lines,
but they are all blended, except for those at 1365.382 and 1385.768\,\AA. 
We adopted the \ion{Au}{ii} and \ion{Au}{iii} line lists from
Rosberg \& Wyart (1997) and Wyart et al. (1996), respectively.\\ 
{\bf Mercury (Hg) Z=80}: 
 The $\log\,gf$ values for \ion{Hg}{i} and \ion{Hg}{ii} were taken from the NIST
database. For \ion{Hg}{iii} we considered only the three lines at 1360.509,
1647.482, and 1738.540\,\AA\ listed by 
Profitt et al. (1999).
The abundance of $-$6.60\,dex  is well suited to reproducing the lines in the ultraviolet,
while $-$6.30\,dex derived from the \ion{Hg}{ii} line at 3984\,\AA\ (Castelli \& Hubrig 2004)
is too large. We did not consider any hyperfine structure in
the computation of the ultraviolet lines.
For \ion{Hg}{i}, the line at 1849.499\,\AA\ is blended, while the line at 2536.521\,\AA\ is very weak.
No other lines with available $\log\,gf$ values were observed. 
Numerous lines of \ion{Hg}{ii} were observed and computed. For instance, the line at
1942.273\,\AA\ is strong, but it is blended with \ion{Mn}{ii} at 1942.344\,\AA. 
Two lines at 
1321.712\,\AA\ and 1331.738\,\AA\ are single and well reproduced, those at 1354.289, 1539.142,
and 1869.226\,\AA\ are strong, but blended.
For \ion{Hg}{iii}, the  line at 1360.509\,\AA\ is computed as too strong,
that at 1647.471\,\AA\  would be adequately
reproduced if the \ion{Mn}{iii} line at 1647.497\,\AA\ were omitted;
finally, the line at 1738.540\,\AA\ is blended with numerous other components.

\Online

\begin{table*}[]
\begin{flushleft}
\caption{New energy levels  of \ion{Mn}{ii} and the predicted lines  
observed in the spectrum. } 
\begin{tabular}{rrcllllllllll}
\noalign{\smallskip}\hline
\multicolumn{4}{c}{Upper level}&
\multicolumn{4}{c}{Lower level}&
\multicolumn{1}{c}{$\lambda$(Ritz)}&
\multicolumn{1}{c}{$log\,gf$}&
\multicolumn{1}{c}{$\lambda$(NIST)}&
\multicolumn{1}{c}{Notes}
\\
\noalign{\smallskip}\hline
\multicolumn{1}{c}{Eup(cm$^{-1}$)}&
\multicolumn{1}{c}{Conf.} &
\multicolumn{1}{c}{Term} &
\multicolumn{1}{c}{J} &
\multicolumn{1}{c}{Elow(cm$^{-1}$)}&
\multicolumn{1}{c}{Conf.} &
\multicolumn{1}{c}{Term} &
\multicolumn{1}{c}{J} &
\multicolumn{1}{c}{\AA} &
\multicolumn{1}{c}{K14} &
\multicolumn{1}{c}{\AA} 
\\
\noalign{\smallskip}\hline
82303.415 & 3d$^{4}$($^{5}$D)4s4p($^{3}$P)& $^{7}$F$^{o}$ & 3.0 & 14325.866& 3d$^{6}$           & a$^{5}$D &  4.0 & 1471.074& $-$3.902\\
          &                               &               &     & 14593.835& 3d$^{6}$           & a$^{5}$D &  3.0 & 1476.896& $-$3.407\\  
          &                               &               &     & 14781.205& 3d$^{6}$           & a$^{5}$D &  2.0 & 1480.994& $-$1.901\\
          &                               &               &     & 29869.420& 3d$^{6}$           & a$^{3}$P &  2.0 & 1907.160& $-$3.174\\
          &                               &               &     & 43395.395& 3d$^{5}$($^{2}$D)4s& c$^{3}$D &  3.0 & 2569.394& $-$0.806\\
          &                               &               &     & 43696.217& 3d$^{5}$($^{4}$F)4s& a$^{5}$F &  3.0 & 2589.416& $-$2.294 & 2589.42 & unassigned\\
\end{tabular}
\end{flushleft}
\end{table*}

\setcounter{table}{0}

\begin{table*}[]
\begin{flushleft}
\caption{Cont.}  
\begin{tabular}{rrcllllllllll}
\noalign{\smallskip}\hline
\multicolumn{4}{c}{Upper level}&
\multicolumn{4}{c}{Lower level}&
\multicolumn{1}{c}{$\lambda$(Ritz)}&
\multicolumn{1}{c}{$log\,gf$}&
\multicolumn{1}{c}{$\lambda$(NIST)}&
\multicolumn{1}{c}{Notes}
\\
\noalign{\smallskip}\hline
\multicolumn{1}{c}{Eup(cm$^{-1}$)}&
\multicolumn{1}{c}{Conf.} &
\multicolumn{1}{c}{Term} &
\multicolumn{1}{c}{J} &
\multicolumn{1}{c}{Elow(cm$^{-1}$)}&
\multicolumn{1}{c}{Conf.} &
\multicolumn{1}{c}{Term} &
\multicolumn{1}{c}{J} &
\multicolumn{1}{c}{\AA} &
\multicolumn{1}{c}{K14} &
\multicolumn{1}{c}{\AA} 
\\
\noalign{\smallskip}\hline
106324.758 &3d$^{6}$($^{4}$G)4d  &$^{5}$F& 4.0 & 64494.140 & 3d$^{5}$($^{4}$G)4p& z$^{5}$G$^{o}$ & 4.0 & 2389.865 & $-$0.611\\
           &                     &       &     & 64518.890 & 3d$^{5}$($^{4}$G)4p& z$^{5}$G$^{o}$ & 5.0 & 2391.280 & $+$0.178 & 2391.281 & unassigned\\
           &                     &       &     & 65658.650 & 3d$^{5}$($^{4}$G)4p& z$^{5}$H$^{o}$ & 5.0 & 2458.306 & $-$0.937 &\\
           &                     &       &     & 66542.539 & 3d$^{5}$($^{4}$G)4p& z$^{5}$F$^{o}$ & 5.0 & 2512.929 & $-$0.462 & 2512.932 & unassigned\\
           &                     &       &     & 66643.296 & 3d$^{5}$($^{4}$G)4p& z$^{5}$F$^{o}$ & 4.0 & 2519.310 & $+$0.320 & 2519.312 & DIFFERENT assign.\\
           &                     &       &     & 66686.739 & 3d$^{5}$($^{4}$G)4p& z$^{5}$F$^{o}$ & 3.0 & 2522.072 & $-$0.278 &\\
\end{tabular}
\end{flushleft}
\end{table*}

\setcounter{table}{0}

\begin{table*}[]
\begin{flushleft}
\caption{Cont.}  
\begin{tabular}{rrcllllllllll}
\noalign{\smallskip}\hline
\multicolumn{4}{c}{Upper level}&
\multicolumn{4}{c}{Lower level}&
\multicolumn{1}{c}{$\lambda$(Ritz)}&
\multicolumn{1}{c}{$log\,gf$}&
\multicolumn{1}{c}{$\lambda$(NIST)}&
\multicolumn{1}{c}{Notes}
\\
\noalign{\smallskip}\hline
\multicolumn{1}{c}{Eup(cm$^{-1}$)}&
\multicolumn{1}{c}{Conf.} &
\multicolumn{1}{c}{Term} &
\multicolumn{1}{c}{J} &
\multicolumn{1}{c}{Elow(cm$^{-1}$)}&
\multicolumn{1}{c}{Conf.} &
\multicolumn{1}{c}{Term} &
\multicolumn{1}{c}{J} &
\multicolumn{1}{c}{\AA} &
\multicolumn{1}{c}{K14} &
\multicolumn{1}{c}{\AA} 
\\
\noalign{\smallskip}\hline
\\
107155.877 & 3d$^{5}$($^{4}$G)4d & $^{3}$I& 5.0& 67846.240 & 3d$^{5}$($^{4}$G)4p& z$^{3}$H$^{o}$ & 5.0 & 2543.142 & $-$0.656 &\\
           &                     &        &    & 67910.560 & 3d$^{5}$($^{4}$G)4p& z$^{3}$H$^{o}$ & 4.0 & 2547.310 & $+$0.810 & 2547.31 & unassigned\\ 
107145 038 & 3d$^{5}$($^{4}$G)4d & $^{3}$I& 6.0& 67744.440 & 3d$^{5}$($^{4}$G)4p& z$^{3}$H$^{o}$ & 6.0 & 2537.27  & $-$0.644 & 2537.265& unassigned \\
           &                     &        &    & 67846.240 & 3d$^{5}$($^{4}$G)4p& z$^{3}$H$^{o}$ & 5.0 & 2543.843 & $+$0.885 & 2543.851 &unassigned\\
107114.165 & 3d$^{5}$($^{4}$G)4d & $^{3}$I& 7.0& 64550.040 & 3d$^{5}$($^{4}$G)4p& z$^{5}$G$^{o}$ & 6.0 & 2348.677 & $-$1.321 &\\
\end{tabular}
\end{flushleft}
\end{table*}

\setcounter{table}{0}

\begin{table*}[]
\begin{flushleft}
\caption{Cont.}  
\begin{tabular}{rrcllllllllll}
\noalign{\smallskip}\hline
\multicolumn{4}{c}{Upper level}&
\multicolumn{4}{c}{Lower level}&
\multicolumn{1}{c}{$\lambda$(Ritz)}&
\multicolumn{1}{c}{$log\,gf$}&
\multicolumn{1}{c}{$\lambda$(NIST)}&
\multicolumn{1}{c}{Notes}
\\
\noalign{\smallskip}\hline
\multicolumn{1}{c}{Eup(cm$^{-1}$)}&
\multicolumn{1}{c}{Conf.} &
\multicolumn{1}{c}{Term} &
\multicolumn{1}{c}{J} &
\multicolumn{1}{c}{Elow(cm$^{-1}$)}&
\multicolumn{1}{c}{Conf.} &
\multicolumn{1}{c}{Term} &
\multicolumn{1}{c}{J} &
\multicolumn{1}{c}{\AA} &
\multicolumn{1}{c}{K14} &
\multicolumn{1}{c}{\AA} 
\\
\noalign{\smallskip}\hline
108521.377 & 3d$^{6}$($^{4}$P)4d &$^{5}$F& 2.0& 66645.116 & 3d$^{5}$($^{4}$P)4p& z$^{5}$D$^{o}$ & 1.0& 2387.260 & $+$0.036 & 2387.268& unassigned\\
           &                &        &        & 66676.833 & 3d$^{5}$($^{4}$P)4p& z$^{5}$D$^{o}$ & 2.0& 2389.070 & $-$0.152 & 2389.075& unassigned\\
           &                &        &        & 66894.130 & 3d$^{5}$($^{4}$G)4p& z$^{5}$F$^{o}$ & 1.0& 2401.542 & $-$0.773 &\\
           &                &        &        & 66901.494 & 3d$^{5}$($^{4}$G)4p& z$^{5}$F$^{o}$ & 2.0& 2401.967 & $-$0.278 &\\
           &                &        &        & 67009.217 & 3d$^{5}$($^{4}$P)4p& z$^{5}$D$^{o}$ & 3.0& 2408.200 & $-$0.818 &\\
           &                &        &        & 72247.730 & 3d$^{5}$($^{4}$D)4p& y$^{5}$D$^{o}$ & 3.0& 2756.008 & $-$1.771 & 2756.01 & unassigned\\
\end{tabular}
\end{flushleft}
\end{table*}

\setcounter{table}{0}

\begin{table*}[]
\begin{flushleft}
\caption{Cont.} 
\begin{tabular}{rrcllllllllll}
\noalign{\smallskip}\hline
\multicolumn{4}{c}{Upper level}&
\multicolumn{4}{c}{Lower level}&
\multicolumn{1}{c}{$\lambda$(Ritz)}&
\multicolumn{1}{c}{$log\,gf$}&
\multicolumn{1}{c}{$\lambda$(NIST)}&
\multicolumn{1}{c}{Notes}
\\
\noalign{\smallskip}\hline
\multicolumn{1}{c}{Eup(cm$^{-1}$)}&
\multicolumn{1}{c}{Conf.} &
\multicolumn{1}{c}{Term} &
\multicolumn{1}{c}{J} &
\multicolumn{1}{c}{Elow(cm$^{-1}$)}&
\multicolumn{1}{c}{Conf.} &
\multicolumn{1}{c}{Term} &
\multicolumn{1}{c}{J} &
\multicolumn{1}{c}{\AA} &
\multicolumn{1}{c}{K14} &
\multicolumn{1}{c}{\AA} 
\\
\noalign{\smallskip}\hline
110611.014 &3d$^{5}$($^{4}$P)4d &$^{3}$F& 3.0& 67812.087 & 3d$^{5}$($^{4}$G)4p& z$^{3}$F$^{o}$ & 3.0  & 2335.791 & $-$0.884&\\
           &                    &       &    & 70231.468 & 3d$^{5}$($^{4}$D)4p& y$^{5}$F$^{o}$ & 2.0  & 2475.753 & $-$0.248& 2475.751 & unassigned\\
           &                    &       &    & 70745.387 & 3d$^{5}$($^{4}$P)4p& z$^{3}$D$^{o}$ & 3.0  & 2507.671 & $-$0.620&\\
           &                    &       &    & 70940.568 & 3d$^{5}$($^{4}$P)4p& z$^{3}$D$^{o}$ & 2.0  & 2520.010 & $+$0.412& 2520.011 & unassigned\\
\\
110998.198 &3d$^{5}$($^{4}$P)4d &$^{3}$P& 2.0& 69044.910 & 3d$^{5}$($^{4}$P)4p& z$^{3}$P$^{o}$ & 2.0  & 2382.877 & $-$0.094 & 2382.88 & DIFFERENT assign.\\
\end{tabular}
\end{flushleft}
\end{table*}

\setcounter{table}{0}

\begin{table*}[]
\begin{flushleft}
\caption{Cont.} 
\begin{tabular}{rrcllllllllll}
\noalign{\smallskip}\hline
\multicolumn{4}{c}{Upper level}&
\multicolumn{4}{c}{Lower level}&
\multicolumn{1}{c}{$\lambda$(Ritz)}&
\multicolumn{1}{c}{$log\,gf$}&
\multicolumn{1}{c}{$\lambda$(NIST)}&
\multicolumn{1}{c}{Notes}
\\
\noalign{\smallskip}\hline
\multicolumn{1}{c}{Eup(cm$^{-1}$)}&
\multicolumn{1}{c}{Conf.} &
\multicolumn{1}{c}{Term} &
\multicolumn{1}{c}{J} &
\multicolumn{1}{c}{Elow(cm$^{-1}$)}&
\multicolumn{1}{c}{Conf.} &
\multicolumn{1}{c}{Term} &
\multicolumn{1}{c}{J} &
\multicolumn{1}{c}{\AA} &
\multicolumn{1}{c}{K14} &
\multicolumn{1}{c}{\AA} 
\\
\noalign{\smallskip}\hline
112106.537& 3d$^{5}$($^{4}$D)4d &$^{5}$F & 3.0 & 70231.468 & 3d$^{5}$($^{4}$D)4p& y$^{5}$F$^{o}$ & 2.0 & 2387.328 & $-$0.921 &\\
          &                     &        &     & 70342.936 & 3d$^{5}$($^{4}$D)4p& y$^{5}$F$^{o}$ & 3.0 & 2393.700 & $-$0.202 & 2393.70 & unassigned.\\
          &                     &        &     & 70497.790 & 3d$^{5}$($^{4}$D)4p& y$^{5}$F$^{o}$ & 4.0 & 2402.610 & $-$0.596 &\\
          &                     &        &     & 70745.387 & 3d$^{5}$($^{4}$P)4p& z$^{3}$D$^{o}$ & 3.0 & 2416.993 & $-$0.679 &\\
          &                     &        &     & 71323.600 & 3d$^{5}$($^{4}$D)4p& x$^{5}$P$^{o}$ & 2.0 & 2451.264 & $-$0.746 &\\
          &                     &        &     & 72247.730 & 3d$^{5}$($^{4}$D)4p& y$^{5}$P$^{o}$ & 3.0 & 2508.100 & $-$0.171 & 2508.102 & unassigned\\
\end{tabular}
\end{flushleft}
\end{table*}

\begin{table*}[]
\begin{flushleft}
\caption{Lines from the NIST database in the 2400$-$3040\,\AA\ interval which are
observed in the spectrum, but are either not predicted or predicted too weak.} 
\begin{tabular}{rrrlllrlllll}
\noalign{\smallskip}\hline
\multicolumn{1}{c}{$\lambda$(NIST)}&
\multicolumn{1}{c}{$log\,gf$}&
\multicolumn{1}{c}{Elow}&
\multicolumn{1}{c}{J} &
\multicolumn{1}{c}{Config.}&
\multicolumn{1}{c}{Term}&
\multicolumn{1}{c}{Eup}&
\multicolumn{1}{c}{J}&
\multicolumn{1}{c}{Config.}&
\multicolumn{1}{c}{Term}&
\multicolumn{1}{l}{Notes}
\\
\noalign{\smallskip}\hline
\multicolumn{1}{c}{$\AA$}&
\multicolumn{1}{c}{}&
\multicolumn{1}{c}{cm$^{-1}$}&
\multicolumn{1}{c}{ } &
\multicolumn{1}{c}{ }&
\multicolumn{1}{c}{ }&
\multicolumn{1}{c}{cm$^{-1}$}&
\multicolumn{1}{c}{}&
\multicolumn{1}{c}{}&
\multicolumn{1}{c}{}&
\multicolumn{1}{l}{}
\\
\noalign{\smallskip}\hline
 2409.045 & $-$3.275 & 55116.333 &  2 &3d$^{4}$4s$^{2}$& c$^{5}$D &  96613.930 & 2 &3d$^{4}$($^{5}$D4s4p($^{3}$P$^{o}$)& s$^{3}$F$^{o}$ & obs\\ 
 2421.089 & $-$3.595 & 49291.360 &  3 &3d$^{6}$        & b$^{1}$F &  90582.550 & 2 & 3d$^{4}$($^{5}$D4s4p($^{3}$P$^{o}$)& w$^{5}$F$^{o}$ & obs at 2421.10 \\
 2431.323 & $-$6.436 & 49465.150 &  3 &3d$^{5}$($^{2}$G)4s & c$^{3}$G &90582.550 & 2 & d$^{4}$($^{5}$D4s4p($^{3}$P$^{o}$)& w$^{5}$F$^{o}$ & obs at 2431.33\\
 2441.226 & $-$5.624 & 73395.444 &  3 &3d$^{5}$($^{4}$D)4p & y$^{3}$D$^{o}$& 114346.060&  3 &3d$^{5}$($^{6}$S)7d & $^{7}$D & obs\\
 2443.392 & $-$5.162 & 35004.790 & 3  &3d$^{6}$            &b$^{3}$G &75919.107& 2 &3d$^{6}$($^{4}$D)4p & y$^{3}$P$^{o}$& obs at 2443.41\\
 2443.809 & $-$4.828 & 27589.360 & 2  &3d$^{5}$($^{4}$G)4s & a$^{5}$G  & 68496.694 & 1 & 3d$^{5}$($^{4}$P)4p & y$^{5}$P$^{o}$& obs blend\\
\end{tabular}
\end{flushleft}
\end{table*}


\begin{thebibliography}{}




\bibitem[Adelman(1994)]{1994MNRAS.266...97A} Adelman, S.~J.\ 1994, \mnras, 
266, 97 

\bibitem[Ansbacher et al.(1985)]{1985CaJPh..63.1330A} Ansbacher, W., 
Pinnington, E.~H., Bahr, J.~L., 
\& Kernahan, J.~A.\ 1985, Canadian Journal of Physics, 63, 1330 

\bibitem[Ayres(2010)]{2010ApJS..187..149A} Ayres, T.~R.\ 2010, \apjs, 187, 
149 


\bibitem[Ayres(2014)]{2014 http} Ayres, T.~R.\ 2014, http://casa.colorado.edu/$\sim$ayres/ASTRAL/ 
 

\bibitem[B{\"a}ckstr{\"o}m et al.(2013)]{2013JPhB...46t5001B} 
B{\"a}ckstr{\"o}m, E., Nilsson, H., Engstr{\"o}m, L., Hartman, H., 
\& Mannervik, S.\ 2013, Journal of Physics B Atomic Molecular Physics, 46, 205001 
\bibitem[Bidelman(1962)]{1962AJ.....67R.111B} Bidelman, W.~P.\ 1962, \aj, 
67, 111 


\bibitem[Bi{\'e}mont et al.(2011)]{2011MNRAS.414.3350B} Bi{\'e}mont, 
{\'E}., Blagoev, K., Engstr{\"o}m, L., et al.\ 2011, \mnras, 414, 3350 


\bibitem[Bi{\'e}mont et al.(2001)]{2001JPhB...34.1869B} Bi{\'e}mont, E., 
Garnir, H.~P., Li, Z.~S., et al.\ 2001, Journal of Physics B Atomic 
Molecular Physics, 34, 1869 


\bibitem[Carpenter et al.(2014)]{2014arXiv1411.1419C} Carpenter, K.~G., 
Ayres, T.~R., \& the ASTRAL Science Teams 2014, arXiv:1411.1419 


\bibitem[Castelli 
\& Bonifacio(1990)]{1990A&AS...84..259C} Castelli, F., \& Bonifacio, P.\ 1990, \aaps, 84, 259 


\bibitem[Castelli et 
al.(1985)]{1985A&AS...59....1C} Castelli, F., Cornachin, M., Morossi, C., \& Hack, M.\ 1985, \aaps, 59, 1 


\bibitem[Castelli 
\& Hubrig(2004)]{2004A&A...425..263C} Castelli, F., \& Hubrig, S.\ 2004, \aap, 425, 263 

\bibitem[Castelli et al.(2008)]{2008JPhCS.130a2003C} Castelli, F., 
Johansson, S., 
\& Hubrig, S.\ 2008, Journal of Physics Conference Series, 130, 012003 

\bibitem[Castelli 
\& Kurucz(2010)]{2010A&A...520A..57C} Castelli, F., \& Kurucz, R.~L.\ 2010, \aap, 520, AA57 

\bibitem[Castelli et 
al.(2009)]{2009A&A...508..401C} Castelli, F., Kurucz, R.~L., \& Hubrig, S.\ 2009, \aap, 508, 401 

\bibitem[Castelli 
\& Parthasarathy(1995)]{1995ASPC...78..151C} Castelli, F., \& Parthasarathy, M.\ 1995, Astrophysical Applications of Powerful New Databases,
A.S.P. Conf. Ser. 78, 151 


\bibitem[Corliss 
\& Bozman(1962)]{1962etps.book.....C} Corliss, C.~H., \& Bozman, W.~R.\ 1962, NBS Monograph, Washington: US Department of Commerce, National Bureau of Standards, |c1962,  

\bibitem[1981]{Cowan1981}
Cowan, R.~D.\ 1981,
The Theory of Atomic Structure and Spectra (Berkeley: Univ. California Press)

\bibitem[Cowley 
\& Hensberge(1981)]{1981ApJ...244..252C} Cowley, C.~R., \& Hensberge, H.\ 1981, \apj, 244, 252 

\bibitem[Den Hartog et al.(2011)]{2011ApJS..194...35D} Den Hartog, E.~A., 
Lawler, J.~E., Sobeck, J.~S., Sneden, C., 
\& Cowan, J.~J.\ 2011, \apjs, 194, 35 


\bibitem[Dworetsky(1969)]{1969ApJ...156L.101D} Dworetsky, M.~M.\ 1969, 
\apjl, 156, L101 


\bibitem[Holt et al.(1999)]{1999MNRAS.306..107H} Holt, R.~A., Scholl, 
T.~J., \& Rosner, S.~D.\ 1999, \mnras, 306, 107 

\bibitem[Iglesias 
\& Velasco(1964)]{1964smi..book.....I} Iglesias, L., \& Velasco, R.\ 1964, The spectrum of the Mn+ ion , by Iglesias, Laura.; Velasco, R.~ Madrid : Consejo Superior de Investigaciones Cientificas, [1964],  


\bibitem[Ivarsson et 
al.(2004)]{2004A&A...425..353I} Ivarsson, S., Wahlgren, G.~M., Dai, Z., Lundberg, H., \& Leckrone, D.~S.\ 2004, \aap, 425, 353 


\bibitem[Kling 
\& Griesmann(2000)]{2000ApJ...531.1173K} Kling, R., \& Griesmann, U.\ 2000, \apj, 531, 1173 

\bibitem[Kling et al.(2001)]{2001ApJS..134..173K} Kling, R., Schnabel, R., 
\& Griesmann, U.\ 2001, \apjs, 134, 173 


\bibitem[Kramida 
\& Sansonetti(2013)]{2013ApJS..205...14K} Kramida, A., \& Sansonetti, J.~E.\ 2013, \apjs, 205, 14 

\bibitem[Kramida et 
al. (2014)] {kra} Kramida, A., Ralchenko, Y., Reader, J., \&  NIST ASD Team \ 2014,
NIST Atomic Spectra Database (ver. 5.2),  available at: http://physics.nist.gov/asd  




\bibitem[Kurucz(1988)]{1988TRANS...IAU} Kurucz, R.~L.\ 1988, Trans. IAU,XXB, edited by McNally
(Dordrecht:Kluwer,1988),pp.168-172. Data available at http://kurucz.harvard.edu/linelists/gfall/


\bibitem[Kurucz(1993)]{1993KurCD..18.....K} Kurucz, R.\ 1993, SYNTHE 
Spectrum Synthesis Programs and Line Data.~Kurucz CD-ROM No.~18.~Cambridge, 
Mass.: Smithsonian Astrophysical Observatory, 1993., 18.
 Data available at http://kurucz.harvard.edu/linelists/gfall/ 


\bibitem[Kurucz(2005)]{2005MSAIS...8...14K} Kurucz, R.~L.\ 2005, Memorie 
della Societa Astronomica Italiana Supplementi, 8, 14 


\bibitem[Kurucz(2011)]{2011CaJPh..89..417K} Kurucz, R.~L.\ 2011, Canadian 
Journal of Physics, 89, 417. Data available at  http://kurucz.harvard.edu/atoms/

\bibitem[Kurucz(2014)]{2014Kur}Kurucz, R.~L.\ 2014, File hyper250155.pos available at 
http://kurucz.harvard.edu/atoms/2501/


\bibitem[Lundberg et al.(1996)]{1996ApJ...469..388L} Lundberg, H., 
Johansson, S.~G., Larsson, J., et al.\ 1996, \apj, 469, 388 


\bibitem[Moore(1950)]{1950aumt.book.....M} Moore, C.~E.\ 1950, NBS 
Circular 488, Washington: US Government Printing Office (USGPO)  

\bibitem[Nave 
\& Johansson(2013)]{2013ApJS..204....1N} Nave, G., \& Johansson, S.\ 2013, \apjs, 204, 1 

\bibitem[Nielsen et al.(2005)]{2005AJ....130.2312N} Nielsen, K.~E., 
Wahlgren, G.~M., Proffitt, C.~R., Leckrone, D.~S., 
\& Adelman, S.~J.\ 2005, \aj, 130, 2312 


\bibitem[Peterson 
\& Kurucz(2015)]{2015ApJS..216....1P} Peterson, R.~C., \& Kurucz, R.~L.\ 2015, \apjs, 216, 1 


\bibitem[Proffitt et al.(1999)]{1999ApJ...512..942P} Proffitt, C.~R., 
Brage, T., Leckrone, D.~S., et al.\ 1999, \apj, 512, 942 


\bibitem[Quinet(1996)]{1996PhyS...54..483Q} Quinet, P.\ 1996, \physscr, 54, 
483 

\bibitem[Quinet et 
al.(2012)]{2012A&A...537A..74Q} Quinet, P., Bi{\'e}mont, E., Palmeri, P., et al.\ 2012, \aap, 537, AA74 




\bibitem[Rosberg 
\& Wyart(1997)]{1997PhyS...55..690R} Rosberg, M., \& Wyart, J.-F.\ 1997, \physscr, 55, 690 


\bibitem[Ryabchikova et al.(1997)]{1997BaltA...6..244R} Ryabchikova, T.~A., 
Piskunov, N.~E., Kupka, F., \& Weiss, W.~W.\ 1997, Baltic Astronomy, 6, 244,
http://vald.inasan.ru/$\sim$vald3/php/vald.php 


\bibitem[Ryabtsev et al.(1993)]{1993PhyS...47...45R} Ryabtsev, A.~N., 
Wyart, J.-F., Joshi, Y.~N., Raassen, A.~J.~J., 
\& Uylings, P.~H.~M.\ 1993, \physscr, 47, 45 




\bibitem[Sugar 
\& Corliss(1985)]{1985aeli.book.....S} Sugar, J., \& Corliss, C.\ 1985, Washington: American Chemical Society, 1985,  


\bibitem[Uylings 
\& Raassen(1997)]{1997A&AS..125..539U} Uylings, P.~H.~M., \& Raassen, A.~J.~J.\ 1997, \aaps, 125, 539 


\bibitem[Warner 
\& Kirkpatrick(1969)]{1969MNRAS.142..265W} Warner, B., \& Kirkpatrick, R.~C.\ 1969, \mnras, 142, 265 


\bibitem[Wyart et al.(1995)]{1995PhyS...52..535W} Wyart, J.-F., Blaise, J., 
\& Joshi, Y.~N.\ 1995, \physscr, 52, 535 


\bibitem[Wyart et al.(1996)]{1996PhyS...53..174W} Wyart, J.-F., Joshi, 
Y.~N., Tchang-Brillet, L., \& Raassen, A.~J.~J.\ 1996, \physscr, 53, 174 





\end{thebibliography}
\end{document}